\documentclass{ws-procs9x6}

\newcommand{\be}{\begin{eqnarray}}
\newcommand{\ee}{\end{eqnarray}}

\newcommand{\bea}{\begin{eqnarray}}
\newcommand{\eea}{\end{eqnarray}}
\newcommand{\Xmax}{X_{\max}}
\newcommand{\Nmax}{N_{\max}}
\newcommand{\ewidth}{0.49\columnwidth}

\newcommand{\sibyll}{\textsc{sibyll~2.1}}
\newcommand{\qgsjet}{\textsc{qgsjet01}}
\newcommand{\seneca}{\textsc{seneca}}
\def\gsim{ \,\, \vcenter{\hbox{$\buildrel{\displaystyle >}\over\sim$}}
 \,\,}
\def\lton{ \,\, \vcenter{\hbox{$\buildrel{\displaystyle <}\over\sim$}}
 \,\,}

\begin{document}

\title{High Energy Cosmic Ray Air Showers and small-$x$ QCD}

\author{H.J.~Drescher, A.~Dumitru}

\address{ Institut f\"ur Theoretische Physik\\
Johann Wolfgang Goethe-Universit\"at\\ Postfach 11 19 32 \\ 
60054 Frankfurt, Germany }

\author{ M.~Strikman }

\address{Department of Physics \\ Pennsylvania State University \\
                University Park, PA 16802, USA}  

\maketitle

\abstracts{ We discuss that hadron-induced atmospheric air showers
from ultra-high energy cosmic rays are sensitive to QCD interactions
at very small momentum fractions $x$ where nonlinear effects should
become important. The leading partons from the projectile acquire
large random transverse momenta as they pass through the strong field
of the target nucleus, which breaks up their coherence. This leads to
a steeper $x_F$-distribution of leading hadrons as compared to low
energy collisions, which in turn reduces the position of the shower
maximum $\Xmax$. We argue that high-energy hadronic interaction models
should account for this effect, caused by the approach to the
black-body limit, which may shift fits of the composition of
the cosmic ray spectrum near the GZK cutoff towards lighter elements.
We further show that present data on $\Xmax(E)$ exclude that the rapid
$\sim 1/x^{0.3}$ growth of the saturation boundary (which is
compatible with RHIC and HERA data) persists up to GZK cutoff
energies.  Measurements of $pA$ collisions at LHC could further test
the small-$x$ regime and advance our understanding of high density QCD
significantly.  }

\section{Introduction}

Today, quite little is known about the origin, the spectrum and the
composition of the highest energy cosmic rays. For example,
AGASA\cite{agasa} found about 10 events with $E>10^{11}$~GeV,  well above the
Greisen-Zatsepin-Kuzmin (GZK) cutoff, $E_{\rm GZK}\simeq 6\cdot10^{10}$~GeV,
 which arises because of
interaction of protons with the cosmic microwave background.
On the other hand, the results of the HIRES\cite{hires} collaboration
agree with the existence of the GZK cutoff (assuming isotropic sources).
Forthcoming Auger\cite{auger} data near the GZK cutoff will provide
higher statistics and hopefully help to resolve this puzzle.

The precise knowledge of the primary cosmic ray properties, that is
the particle type, energy and arrival direction, is
crucial for the interpretation of their possible source and
acceleration mechanism. Standard candidates for the highest energy
cosmic rays are protons or heavier nuclei, being accelerated in
extreme astrophysical phenomena, or photons, arising for example
from the decay of ultra-heavy $X-$particles. 

Experiments detect cosmic rays indirectly via air showers induced
when they enter the atmosphere. One tries to deduce the
properties of the primary particle from those of the induced shower.
Therefore, a good understanding of
the physics of high-energy interactions in the atmosphere is
mandatory. However, the maximum energies
exceed those of terrestrial accelerators by far, and so our knowledge
of hadronic interactions needs to be extrapolated to unknown regimes. Also,
as will be discussed in more detail below, air showers are mostly sensitive
to forward particle production which is less well measured in
accelerator experiments.

Several features of strong interactions are expected to change
dramatically at very high energies.
First, the parameters of the soft
interactions change - the total cross section changes by a factor of $\sim
3$, while average impact parameters increase by $\sim50\%$.
The changes in the (semi-)hard interactions are even more
dramatic. Indeed, a leading parton from the projectile propagates through
the very strong gluon fields in the target.  For example, for a
``low-energy'' $p+A$ collision with $E_{\rm Lab}=400$~GeV, a parton
from the projectile carrying a momentum fraction $x_p\sim
0.1 $ receiving a transverse kick of $p_t \sim 2$~GeV
interacts with a gluon with $x_A=4p^2_{\perp}/x_ps\sim 0.1$.
At GZK energy, $E_{\rm Lab}\sim 10^{11}$~GeV, this
corresponds to $x_A\sim 10^{-10}$ while {\it direct measurements} at
HERA covered only the range $x\ge 10^{-3}$ and even indirect ones are
sensitive only down to $x\ge 10^{-4}$.
This is {\it six orders of magnitude above the
$x$-range to which cosmic rays near the cutoff are sensitive.}

Studies at HERA indicate that the gluon density of the nucleon, $xg_N(x,Q^2)$,
grows very strongly with decreasing momentum fraction $x$, roughly as
$x^{-\lambda(Q^2)}$, with $\lambda(Q^2) \ge 0.2$ for $Q^2\ge 2$~GeV$^2$.
The data can be fitted by the NLO QCD evolution equations.  The
analysis of partial waves for the interaction of a small dipole with
the nucleon at
HERA energies indicates that for $q\bar q$ dipoles the partial
waves remain substantially below the unitarity limit, while for $gg$
dipoles the unitarity limit is practically reached for virtualities $Q^2\simeq
4$~GeV$^2$ at top HERA energies.  This indicates that for
higher energies further growth of the gluon density in the proton
should be tamed for a range of
virtualities that increases with energy.

The rate at which the growth of the gluon density is ``tamed'', depends
strongly on its behavior at small $x$.  One of the most
popular approaches for a long time was the BFKL approximation where
one sums the series of leading, next to leading $\log (1/x)$.  The
series was found to converge poorly, to a large extent due to the
specifics of the treatment of the energy conservation effects.  More
recent calculations\cite{Altarelli:2003hk,Ciafaloni:2003rd} (some of
which were discussed at the Erice meeting) try to treat
simultaneously logarithms of both $1/x$ and $Q^2$ and to treat more
accurately the phase space available for gluon emission at a given
energy.  They appear to indicate that the NLO DGLAP approximation should
effectively work for $x\ge 10^{-3}$ for the scattering off a gluon or,
correspondingly, for $x\ge 10^{-4}$ for the scattering off a nucleon,
which is consistent with the HERA findings.
   
It appears natural to expect that the taming effects would still allow
the interactions to reach the maximal possible strength
allowed by unitarity over a wide range of impact parameters which should
increase with energy.  Indeed, this is implemented in all models
currently on the market with the only difference being the rate of the
approach to the unitarity limit.
  
If the energies are large enough, the constituents of the projectile
hadron/photon propagating through the nucleus resolve strong small-$x$ gluon
fields in the target. During the propagation through such media they
should experience strong distortions - at the
very least they should obtain significant transverse momenta inversely
proportional to the size of test dipoles for which the interaction becomes
black. Also, some of the processes relevant in this case, like hard
scattering of the projectile partons off small-$x$ partons, lead to
fractional energy losses.
   
However the most important effect for the purposes of near-GZK
inelastic collisions is loss of coherence of the leading partons of the
projectile as they acquire random transverse momenta.  This leads to
independent fragmentation of the leading partons from the projectile
over a large range of rapidities, and hence to a much softer energy spectrum
of the leading particles\cite{bbl_gammaA,DGS}.

In this paper we review our first efforts to model this effect and to
show its relevance for the understanding of air showers induced by
cosmic rays near the GZK cutoff. Our primary goal is to analyze 
the implication of various
models for the small-$x$ behavior of the gluon densities beyond
the HERA range.
We demonstrate that already the current data on the longitudinal and
lateral structure of giant air showers allow us to
rule out certain models where gluon densities increase very rapidly with
energy\cite{DDS}.  At the same time, models which appear to be consistent with
the recent theoretical studies\cite{Altarelli:2003hk,Ciafaloni:2003rd} lead to
relatively small effects which are consistent with the air shower data and
suggest that the spectrum near the cutoff is dominated by protons.

Finally, we also point out that the gluon densities encountered in
central proton-nucleus collisions at LHC are similar to those for central
proton-air collisions near the cutoff energy. Hence, we also present some
predictions for p-"heavy nucleus" central collisions at RHIC and
LHC energies which could test our suggestion for the mechanism of
energy degradation by projectile breakup. Such measurements could help
us understand the interactions of very high energy cosmic rays with
the atmosphere and, consequently, of their composition and origin.


\section{Scattering at high energies}

The wave function of a hadron (or nucleus) boosted to large rapidity exhibits
a large number of gluons at small $x$. The
density of gluons per unit of transverse area and of rapidity at
saturation is denoted by $Q_s^2$, the so-called saturation momentum. This
provides an intrinsic momentum scale which grows with atomic number (for
nuclei) and with rapidity, due to continued gluon radiation as phase space
grows. For sufficiently high energies and/or large nuclei, $Q_s$
can grow much larger than $\Lambda_{\rm QCD}$ and so weak coupling methods
are applicable. Nevertheless, the well known leading-twist pQCD can not be used
when the gluon density is large; rather,
scattering amplitudes have to be resummed to all orders in the density. When
probed at a scale below $Q_s$, scattering cross sections approach the
geometrical size of the hadron (the ``black body'' limit). 
A perturbative QCD based mechanism for unitarization of cross sections is 
provided by gluon saturation effects~\cite{mq,sat,mueller}.
On the other hand, for
$Q^2\gg Q_s^2$ the process occurs in the dilute DGLAP~\cite{DGLAP} 
regime where cross sections are approximately
determined by the known leading-twist pQCD expressions.

In this section we discuss particle production in the collision of a
hadron, which for the present purposes is either a nucleon or a meson,
with a target nucleus in the atmosphere.  The development of the air
shower is sensitive mainly to the distribution of the most energetic
particles (see section~\ref{sec_AirSh}) while the low-energy particles
produced near the nuclear fragmentation region are less important for
the observables studied here. Due to QCD evolution
(section~\ref{Qs_rap}) this so-called forward region probes
the high gluon density (small-$x$) regime of the target,
while the density of the
projectile is rather low.  Hence, in the relevant rapidity (or
Feynman-$x_F$) region we are dealing with a ``dilute'' projectile
hadron impinging on a ``dense'' target nucleus: $Q_s^h < Q_s^A$.

This assumption breaks
down at large impact parameters, where even the saturation momentum of the
nucleus, $Q_s^A(y,b)$, is not large. For such events, as well as for
collisions at moderately high energies, no intrinsic semi-hard scale
exists in the problem and so a treatment within weak coupling QCD is
not applicable. In accelerator experiments this regime could be
avoided by appropriately tuning the control parameters, such as
collision energy, atomic number of projectile and target, impact
parameter (trigger), rapidity $y$, transverse momentum $p_t$ and so
on. This, of course, is not feasible in the case of cosmic ray air
showers; here, we model such collisions using the SIBYLL leading-twist
event generator. This is discussed in more detail in
section~\ref{sec_BBLMC}.

\subsection{Leading quarks}
With this in mind, we now focus on particle production in collisions
at sufficiently high energy and sufficiently small impact parameter
where the saturation momentum of the nucleus is large enough to
warrant a weak-coupling approach.
The dominant process for fast particle production ($x_F\gsim0.1$) is
scattering of quarks from the incident dilute projectile on the dense target. 
For high quark energy we assume that the eikonal approximation applies
such that $p^+$ is conserved. The transverse momentum distribution of
scattered quarks is then given by the correlation function of two Wilson
lines,
\be
V(x_t) = \hat{P} \exp\left[ -ig \int_{-\infty}^\infty dz^-
  A^+(z^-,x_t) \right]~,
\ee
running up and down the light cone at transverse separation $r_t$
(in the amplitude and its complex conjugate),
\be
\sigma^{qA} = \int \frac{d^2q_t dq^+}{(2\pi)^2} \delta(q^+ - p^+)
\left<\frac{1}{N_c}\,{\rm tr}\,
\left| \int d^2 z_t \, e^{i\vec{q}_t\cdot \vec{z}_t} \left[
V(z_t)-1\right] \right|^2\right>~.
\ee
Here, the convention is that the incident hadron has positive rapidity, i.e.\
the large component of its light-cone momentum is $P^+$, and that of the
incoming quark is $p^+=x P^+$ ($q^+$ for the outgoing quark).
The two-point function has to be evaluated in the background field of the
target nucleus. When this field is weak, the Wilson lines can be
expanded to leading order and the problem reduces to evaluating the
two-point function of the gauge field $A^\mu$.

In the strong-field regime $gA^+\sim1$, however, one needs to evaluate
the correlation function to all orders (corresponding to summing over
any number of scatterings of the incident quark). 
A relatively simple closed expression can be obtained in the
McLerran-Venugopalan model of the small-$x$ gluon distribution of the
dense target\cite{sat}.
In that model, the small-$x$ gluons are described as a stochastic
classical non-abelian Yang-Mills field which is averaged over with a Gaussian
distribution. $n$-point functions then factorize into powers of the
two-point function. The $qA$ cross section is then given by\cite{djm2} 
\bea \label{qAXsec}
q^+ \frac{d\sigma^{qA\to qX}}{dq^+d^2q_t d^2b} &=& \frac{q^+}{P^+} \,
\delta\left(
\frac{p^+ - q^+}{P^+}\right) C(q_t)\\
C(q_t) &=& \int \frac{d^2r_t}{(2\pi)^2} \, e^{i\vec{q}_t\cdot \vec{r}_t}
\left\{
\exp\left[-2Q_s^2 \int_\Lambda \frac{d^2 l_t}{(2\pi)^2}\frac{1}{l_t^4}
\left(1-e^{i \vec{l}_t \cdot \vec{r}_t}\right)\right] \right.\nonumber\\
& & \left. ~~~-2\exp\left[-Q_s^2\int_\Lambda
 \frac{d^2 l_t}{(2\pi)^2}\frac{1}{l_t^4}\right]
+1\right\}~. \label{Cqt}
\eea
This expression is valid to leading order in $\alpha_s$ (tree level), but to 
all orders in $Q_s$ since it resums any number of scatterings of the
quark in the strong field of the nucleus. The saturation momentum $Q_s$, as
introduced in eq.~(\ref{qAXsec}), is related to $\chi$, the total
color charge density squared (per unit area) from the nucleus integrated
up to the rapidity $y$ of the probe (i.e.\ the projectile quark), by
$Q_s^2 = 4\pi^2\alpha_s^2 \; \chi\; (N_c^2-1)/{N_c}$.
In the low-density limit, $\chi$ is proportional to the ordinary leading-twist
gluon distribution function of the nucleus\cite{gy_mcl}:
\be
\chi(x) = \frac{A}{\pi R_A^2} \int_x^1 dx' \left(
\frac{1}{2N_c} \; q(x',Q_s^2) + \frac{N_c}{N_c^2-1} \; g(x',Q_s^2)\right)~,
\ee
where $q(x)$ and $g(x)$ denote the quark and gluon distributions of a
nucleon, respectively; note that shadowing in the linear
regime\cite{FS88} would tend to reduce $\chi$ somewhat but is
neglected here since we are dealing with small nuclei (mass number
$14\to 16$) and because the induced air showers are sensitive mainly
to the small-$x$ regime in the nucleus.

The integrals over $p_t$ in eq.~(\ref{qAXsec}) are
cut off in the infrared by some cutoff $\Lambda$,
which we assume is of order $\Lambda_{\rm QCD}$.
At large transverse momentum, again
the first exponential in~(\ref{qAXsec}) can be expanded order 
by order\cite{gelis,Boer} to generate the usual power series in
$1/q_t^2$:
\be
C(q_t) =\frac{1}{2\pi^2}
 \frac{Q_s^2}{q_t^4}\left[1+\frac{4}{\pi}\frac{Q_s^2}{q_t^2}
\log\frac{q_t}{\Lambda} + {
O}\left(\frac{Q_s^2}{q_t^2}\right)\right]~.
\label{LT}
\ee
This expression is valid to leading logarithmic accuracy. The first term
corresponds to the perturbative one-gluon $t$-channel exchange contribution 
to $qg \to qg$ scattering and exhibits the well-known power-law
divergence of leading-twist perturbation theory for small momentum
transfer. 

On the other hand, for $Q_s\gsim q_t$ one obtains
in the leading logarithmic approximation\cite{Boer}
\be \label{Cqt_Qs}
C(q_t) \simeq \frac{1}{Q_s^2\log\,Q_s/\Lambda}\, \exp\left(-
\frac{\pi q_t^2}{Q_s^2\log\,Q_s/\Lambda}\right)~.
\ee
This approximation reproduces the behavior of the full
expression~(\ref{qAXsec}) about $q_t\sim Q_s$, and hence the
transverse momentum integrated cross section reasonably well.
It is useful when the cutoff
$\Lambda\ll Q_s$, that is, when color neutrality is enforced on
distance scales of order $1/\Lambda\gg1/Q_s$. If, however, color
neutrality in the target nucleus occurs over distances of order
$1/Q_s$\cite{kazu} then $\Lambda\sim Q_s$ and one has to go beyond
the leading-logarithmic approximation. 

It is essential to realize that the high-energy part of the air shower
is essentially one-dimensional, i.e.\ the transverse momenta of the
produced {\em hadrons} play no role\footnote{We repeat, however, that the
  transverse broadening of the distributions of released {\em partons does}
play an important role since it destroys the coherence of the
projectile wave function\cite{DGS} and affects the fragmentation into
hadrons.} (see section~\ref{sec_AirSh}). This, in turn,
implies that when $Q_s$ is large that the high-transverse momentum
leading-twist regime can be neglected. The $q_t$-distribution of
forward valence quarks can thus be taken to be given by the simple
expression~(\ref{Cqt_Qs}) rather than~(\ref{Cqt}). Note also that both
expressions do conserve probability, i.e.\ 
\be
\int d^2q_t~C(q_t) = 1~.
\ee
This is, of course, a very useful property
because all charges carried by the valence quarks are then
automatically conserved.

Contrary to the leading twist expression~(\ref{LT}), the
distribution~(\ref{Cqt_Qs}) exhibits transverse broadening as the
density of the target increases (the scattered quarks are pushed out
to larger $q_t$).
Consider now the probability of inelastic scattering (i.e.\ {\em with}
color exchange) to small transverse momentum. This is given 
by expression~(\ref{Cqt},\ref{Cqt_Qs}), integrated from $q_t=0$ to
 $q_t=\Lambda$:
\be
\int\limits_0^\Lambda d^2q_t~C(q_t) \simeq
\frac{\pi\Lambda^2} {Q_s^2\log\,Q_s/\Lambda} + \cdots~.
\ee
Here, we have written only the leading term in
$\Lambda^2/Q_s^2$, neglecting subleading power-corrections
and exponentially suppressed contributions.
Hence, soft forward inelastic scattering is power-suppressed in the
black body limit because the typical transverse momentum is
proportional to $Q_s^A$. This steepens the longitudinal distribution
$dN/dx_F$ of leading particles since partons with large relative
momenta fragment independently\cite{bbl_gammaA,DGS}. 
On the other
hand, for low target density, the projectile's coherence is not
destroyed completely and leading quarks may recombine, recovering the
``leading-particle'' effect observed in $pp$ scattering
at not too high energy. This recombination effect should be taken into
account when modeling minimum bias $pA$ collisions in order to ensure
a smooth transition from the high-density to the low-density regime;
our implementation is described and studied in more detail in
section~\ref{sec_BBLMC}.

Integrating over the transverse momentum of the scattered quark,
the elastic and total scattering cross sections for quark-nucleus
scattering are\cite{djm2}:
\be
\sigma^{\rm el} &=& \int d^2b \left[ 1-\exp(-Q_s^2/4\pi\Lambda^2)\right]^2\\
\sigma^{\rm tot} &=& 2\int d^2b \left[ 1-\exp(-Q_s^2/4\pi\Lambda^2)\right]~.
\ee
Clearly, when $Q_s/\Lambda\to\infty$, the cross section approaches the
unitarity limit.

\subsection{Gluons}
Gluon bremsstrahlung dominates particle production at $x_F\lton0.1$.
At very large transverse momentum, $q_t\gg Q_s$, the inclusive gluon
distribution is given in collinear factorization by the
usual $gg\to gg$ LO hard scattering function convoluted with the DGLAP
evolved leading-twist gluon distribution of the projectile and target.
However, for the high-energy part of the air shower only the
$p_t$-integrated longitudinal distribution of hadrons matters (cf.\
section~\ref{sec_AirSh}) which is dominated by fragmentation of gluons
with transverse momenta up to $\sim Q_s^A$. In that regime
leading-twist perturbative QCD can not be applied reliably.

Gluon radiation with transverse momentum
$q_t\sim Q_s^A$ in high-energy hadron-nucleus collisions has been
discussed in detail in\cite{glue,KL}. The release of gluons from the
hadronic wave functions can be described by convoluting the gluon
distribution in the hadron with a (semi-)hard scattering cross section.
 The main qualitative features
of the bremsstrahlung spectrum is that it flattens from $\sim1/q_t^4$
for asymptotically large $q_t$ to $~1/q_t^2$ for
$q_t$ in between the two saturation momenta; for $q_t\to0$, finally, it
approaches a constant (up to logarithms)\cite{KrasVenu}.

\begin{figure}[tb]
\begin{center}
\includegraphics[width=\ewidth]{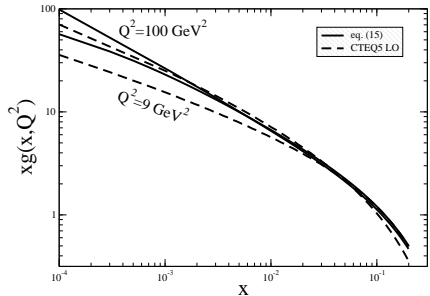}
\end{center}
\caption{Comparison of the gluon distribution from eq.~(\ref{KLNxg})
  with CTEQ5 LO, in the DGLAP regime (for a proton).
 \label{fig:xg}}
\end{figure}

For the present purposes we require a simple {\em ansatz} that can be
easily implemented in a Monte-Carlo model and, at the same time, does
incorporate the above features. A useful approach has been suggested
in refs.\cite{KL}. The ``fusion'' of two gluon ladders gives rise to
the bremsstrahlung spectrum 
\be \label{BrmsSpec}
E\frac{d\sigma}{d^3q} = 4\pi
\frac{N_c}{N_c^2-1} \frac{1}{q_t^2} \int\limits^{q_t^2} d k_t^2 \,
\alpha_s(k_t^2)\, \phi_h(x_1,k_t^2)\, \phi_A(x_2,(q_t-k_t)^2)~, 
\ee
where $\phi(x,Q^2)$ denotes the unintegrated gluon distribution
function of the projectile hadron or target nucleus, respectively.  It
is related to the gluon density by 
\be x\, g(x,Q^2) =
\int\limits^{Q^2} dk_t^2 \, \phi(x,k_t^2)~.  
\ee 
Eq.~(\ref{BrmsSpec}) can be integrated by parts to read
\be\label{SpecInt} 
\frac{dN}{dy dq_t^2} = 
4\pi\alpha_s(q_t^2)\; \frac{N_c}{N_c^2-1} \frac{1}{q_t^4}
\; x_1g_h(x_1,q_t^2)\, x_2g_A(x_2,q_t^2)~. 
\ee
The {\em ansatz}
from\cite{KL} for the infrared-finite gluon densities is 
\be\label{KLNxg} 
x\, g(x,Q^2) \propto \frac{1}{\alpha_s}\, {\rm
min}(Q^2,Q_s^2(x))\,\, (1-x)^4~, 
\ee 
with $\alpha_s$ evaluated at
$\mbox{max}(Q_s^2,Q^2)$. Note that for large $Q^2$, the $x$-dependence
of the gluon distribution exhibits the conventional $xg(x)\sim
x^{-\lambda} (1-x)^4$ behavior; this follows from the 
evolution of the saturation momentum\footnote{For fixed coupling
  evolution this is true for any $x$; for running coupling evolution
  it holds only for not too small $x$, see section~\ref{Qs_rap}.}
$Q_s(x)\sim 1/x^\lambda$ with $x$. On the other hand, for small $Q^2$
and $x$, the above {\em ansatz} exhibits a slow logarithmic growth
$xg(x) \sim \log~x^{-\lambda}$ only. In any case, we
consider~(\ref{KLNxg}) to be a simple parameterization which exhibits
some generic qualitative features expected from gluon saturation
(e.g.\ that it is of order $1/\alpha_s$ at small $Q^2$ and $x$) while,
at the same time, it is roughly consistent with the DGLAP gluon
distribution at large $Q^2$ and $x$. In fig.~\ref{fig:xg} we compare the
parameterization~(\ref{KLNxg}), with the normalization constant fixed
by the condition $\int dx\;xg(x,Q^2)=0.5$ and with $Q_s^2\sim
x^{-0.3}$, to the CTEQ5 LO distribution\cite{cteq5}.

The constant of proportionality in eq.~(\ref{KLNxg})
was chosen in\cite{KL} such as to
reproduce the overall normalization of the charged hadron rapidity
distribution in $d+Au$ collisions at BNL-RHIC energy. This was
possible because the forward region was not considered.  On the other
hand, here we need to consider the entire solid angle (in momentum
space) and, in particular, ensure conservation of the energy carried
by the projectile. In our approach we therefore fix the overall number
of radiated gluons by the condition of energy conservation. This is
discussed in more detail in section~\ref{sec_BBLMC} where we also show
that the charged hadron multiplicity in the central region of $pA$
collisions at BNL-RHIC energy agrees roughly with that from\cite{KL}
and with available data.

\subsection{The saturation momentum as a function of impact parameter
  and rapidity} \label{Qs_rap}

The saturation momentum of the nucleus, $Q_s^A$, must of course depend
on the impact parameter as it basically measures the color charge density in
the transverse plane. Hence, in the rest frame of the nucleus, 
the most naive estimate (neglecting shadowing\cite{FS88}) is that there are
$A$ times more valence quarks in a nucleus than in a nucleon which are
distributed over an area proportional to $A^{2/3}$; this then results
in $(Q_s^A)^2 \sim A^{1/3}$. More elaborate estimates lead to an
additional factor equal to the logarithm of the mass number. 

For realistic nuclei with a non-uniform density distribution in the
transverse plane we must replace, of course, the $A^{1/3}$ factor by
the number of nucleons from the target which interact with the
projectile, $N_A$:
\be \label{Npart}
N_A(b) = A \sigma_{\rm in}(s) T_A(b)
\ee
with $T_A(b)$ the nuclear profile function\footnote{Normalized
  according to $\int d^2b T_A(b)=1$.} and $\sigma_{\rm in}(s)$ the
energy dependent inelastic (non single-diffractive) cross section for
the particular projectile species on protons.
Fluctuations in $N_A(b)$ are also taken into
account in our Monte-Carlo implementation, as discussed in
section~\ref{sec_BBLMC}. The impact parameter dependence of the
nuclear saturation momentum is then taken to be
\be \label{Qs_Np}
Q_s^A(b) = \Lambda \sqrt{\left[1+N_A(b)\right] ~\log (1+N_A(b))}~.
\ee
For a single nucleon, this corresponds to a saturation momentum
on the order of $\Lambda$. It should be noted that at very high
energies, deep in the black-body limit, the results depend only weakly
on the above ``initial condition'' for $Q_s^A$. However, when the
saturation momentum is only moderately large (for example, for running
coupling evolution, see below), the assumed dependence of $Q_s^A$ on
$b$ could play a role. Whether or not $Q_s^2\sim N_A$ is
the most appropriate choice will be studied in more detail in the future.

Next, we turn to the dependence of $Q_s$ on rapidity\footnote{In this
  section, we measure the rapidity always relative to the parent
  hadron.}  $y=\log 1/x$. Eq.~(\ref{Qs_Np}) provides the initial
  condition in (or near) the rest frame of the nucleus, $y\simeq0$,
  from valence quarks. As one moves away in rapidity phase space for
  gluon radiation opens up and so the gluon density grows; it 
is expected to saturate
when it becomes of order $1/\alpha_s$~\cite{mueller}. For a recent
review of evolution at small $x$ see e.g.\cite{IancuVenu}.

Model studies of deep inelastic scattering (DIS) on protons
at HERA suggest\cite{gbw}
\begin{equation} \label{Qs_fc}
Q_s^2(x)\sim x^{-\lambda}
\end{equation}
with $\lambda\approx 0.3$.
This scaling relation can be obtained from the {\em fixed coupling}
BFKL
evolution equation for the scattering amplitude of a small dipole. The
BFKL equation is a linear QCD evolution equation which can not be applied
in the high-density regime. Nevertheless, one can evolve the wave
function of the target in rapidity $y=\log 1/x$ and ask when the
dipole scattering amplitude becomes of order one, which leads
to\footnote{We write only the leading term proportional to $y$.}
\be
Q_s^2(y,b) = Q_s^2(y_0,b) \exp \,c\bar{\alpha}_s y~,
\ee
 with $\bar{\alpha}_s=\alpha_s N_c/\pi$ and
  $c\approx 4.84$ a constant. 
Hence, LO fixed-coupling BFKL evolution predicts $\lambda'=c \bar{\alpha}_s$
of order one, a few times larger than the fit~(\ref{Qs_fc}) to HERA
phenomenology. A resummed NLO BFKL analysis corrects this discrepancy and
leads to $\lambda'$ much closer to
the phenomenological value\cite{Triantafyllopoulos:2002nz}.
A similar observation is made in\cite{Ciafaloni:2003rd} where both
$\log (1/x)$ and $\log Q$ effects were considered.

On the other hand one could also consider BFKL evolution with ad-hoc
one-loop running of the coupling\cite{IancuVenu}\footnote{Again, we
  drop subleading terms that grow more slowly with rapidity.}:
$\bar{\alpha}_s(Q_s^2) =
b_0/\log(Q_s^2(x)/\Lambda_{\rm QCD}^2)$, which leads to
\be \label{Qs_rc}
Q_s^2(y,b) = \Lambda_{\rm QCD}^2 \exp\sqrt{2b_0 c (y+y_0)}~,
\ee
with $2b_0 c y_0 = \log^2 (Q_0(b)^2/\Lambda_{\rm QCD}^2)$.
Insisting that~(\ref{Qs_fc}) be valid at least in the $y\to0$ limit
again provides us with a phenomenological value for the constant $c$
in terms of the saturation momentum at $y=0$.
The form~(\ref{Qs_rc}) leads to
a notably slower growth of $Q_s$ at high energy.
Specifically, for central proton-nitrogen collisions at RHIC, LHC and
GZK-cutoff energies (total rapidity $y=10.7$, 17.3 and 26.0) the
saturation momentum of the nucleus in the rest frame of the projectile
hadron is $Q_s=1.5$, 5, 20~GeV for fixed coupling evolution, while for
running coupling evolution it is $Q_s=1$, 2.5, 6~GeV,
respectively. Clearly, cosmic ray interactions in our atmosphere
should offer a realistic opportunity for distinguishing these scenarios.

\section{Monte-Carlo implementation}
\label{sec_BBLMC}

We first generate a configuration of valence quarks
according to the distribution~(\ref{qAXsec},\ref{Cqt_Qs}), convoluted
with the respective valence quark distribution of the projectile at
the scale $Q_s^A$~\cite{DGS}:
\be \label{valq_distrib}
\frac{d\sigma}{dx d^2q_t d^2b} = f_{\rm v} (x,Q_s^A) ~C(q_t)
\ee
with $Q_s^A$ is a function of both $x$ and $b$.
For this purpose, we employ the GRV94 parameterization of the parton
distribution functions of a proton\cite{GRV94p} or a
$\pi^+$\cite{GRV94pi}; we assume isospin symmetry to deduce the
distributions for other states. Also, as a rough approximation we take
the valence quark distribution of the $K^+$ to be the same as that of
the $\pi^+$, with the replacement $\bar{d}\to\bar{s}$.

The remaining momentum is then used to generate a number of
 gluons according to the distribution~(\ref{SpecInt}).
These gluons could be fragmented independently but this is not a good
approximation when their transverse momenta are soft. Rather, soft collinear
gluons should be absorbed by the parent parton.

The Lund string model\cite{PY} provides such an infrared safe fragmentation
prescription. We order the produced gluons in rapidity and place them
on strings between the valence quarks and the target nucleus, whose
precise configuration is not important. (Target fragmentation produces
only low-energy particles which do not affect the properties of the
air shower discussed here.)

A baryon-nucleus collision produces three strings (two for a
meson-nucleus collision).
However, when the invariant mass of any two of the three
valence quarks is small, one cannot assume anymore that those strings fragment
independently. Rather, they will recombine to form a leading diquark,
recovering the ``leading particle effect'' for low $Q_s^A$.
This effect should be taken into account in order to ensure a smooth
transition from the regime of high target density (high energy, central
collisions) to low target density (lower energy, large impact parameter).
We model this by introducing a cut-off in invariant mass,
\be
m_{\rm cut} = m_{\rho} = 0.77 {\rm~GeV}
\ee
below which  two leading quarks are allowed to form a diquark. The
effect on the $x_F$-distribution of fast hadrons is shown below.

Although soft gluon absorption and diquark recombination are taken
into account in our Monte-Carlo implementation of scattering near the
black body limit of QCD (``BBL''), it should nevertheless be clear that it is
restricted to the high-density regime. For example, when $Q_s^A$
becomes small, the DGLAP leading-twist regime becomes important and
one should use better approximations for the gluon densities than
those from~(\ref{KLNxg}). Also, the fraction of diffractive and
elastic events becomes sizable. 

A large amount of work has been done to develop models for this
regime. SIBYLL\cite{Sibyll} and QGSJET\cite{qgsjet}, in particular,
are commonly used to model air showers. We do not intend to duplicate
those approaches here but rather to study whether anything could be
learned about small-$x$ QCD from cosmic ray air showers. Hence, we
couple our model to the standard pQCD leading twist event generator
\sibyll~such that the ``BBL'' Monte-Carlo treats the high density
regime (large saturation momentum of the nucleus, i.e.\ high energy
and/or small impact parameter) while SIBYLL handles peripheral or low
energy collisions where the saturation momentum of the nucleus is not
sufficiently large. 

It is clear, of course, that no sharp boundary
between those regimes exists and that this artificial separation is
performed for purely technical reasons. It is therefore important to
check that the results do not depend strongly on the precise location
of the assumed boundary between low and high density. As
already mentioned above, we do implement some effects into the BBL
model which should facilitate a smooth transition to low densities,
such as soft gluon absorption and diquark recombination. At the same
time, the SIBYLL model also assists the transition to high target
densities by implementing a low-$p_t$ cutoff for the DGLAP regime
which grows rapidly with energy, see Engel et al\cite{Sibyll}. We therefore
expect that our results are not very sensitive to where exactly we perform the
switch, as long as it occurs in a reasonable regime; this will be
checked below.

The saturation momentum of the nucleus provides an
intrinsic scale for resolving the valence quark structure of the
projectile, cf.\ eq.~(\ref{valq_distrib}). In practice, this scale can
not be too small because the $Q^2$ evolution of parton distributions
in hadrons is normally obtained from DGLAP; standard PDF parameterizations
typically require a minimal $Q^2$ on the order of 1~GeV$^2$.
In order not to distort the inclusive momentum distribution of valence
quarks, we must therefore ensure that $Q_s^A(x)$ does not drop below
this threshold at too large an $x$. Specifically, we
require that the valence quark distribution be probed at least
down to $x=10^{-3}$:
\be \label{Qs_xmin}
Q_s^A(x=10^{-3},b) > Q_{\min} \approx 1~{\rm GeV}  ~.
\ee
In our Monte-Carlo approach, the collision is handled by either BBL or
SIBYLL depending on whether this condition is met or not. 
The resulting boundary between low and high density regimes appears
reasonable; for example, for central collisions of protons on heavy
targets like $Au$ or $Pb$, the transition occurs just below BNL-RHIC
energy, $\sqrt{s}=200$~GeV. On the other hand, minimum bias $pp$
collisions essentially never pass the threshold~(\ref{Qs_xmin}), even
at LHC energies and beyond.

\begin{figure}[tb]
\includegraphics[width=\ewidth]{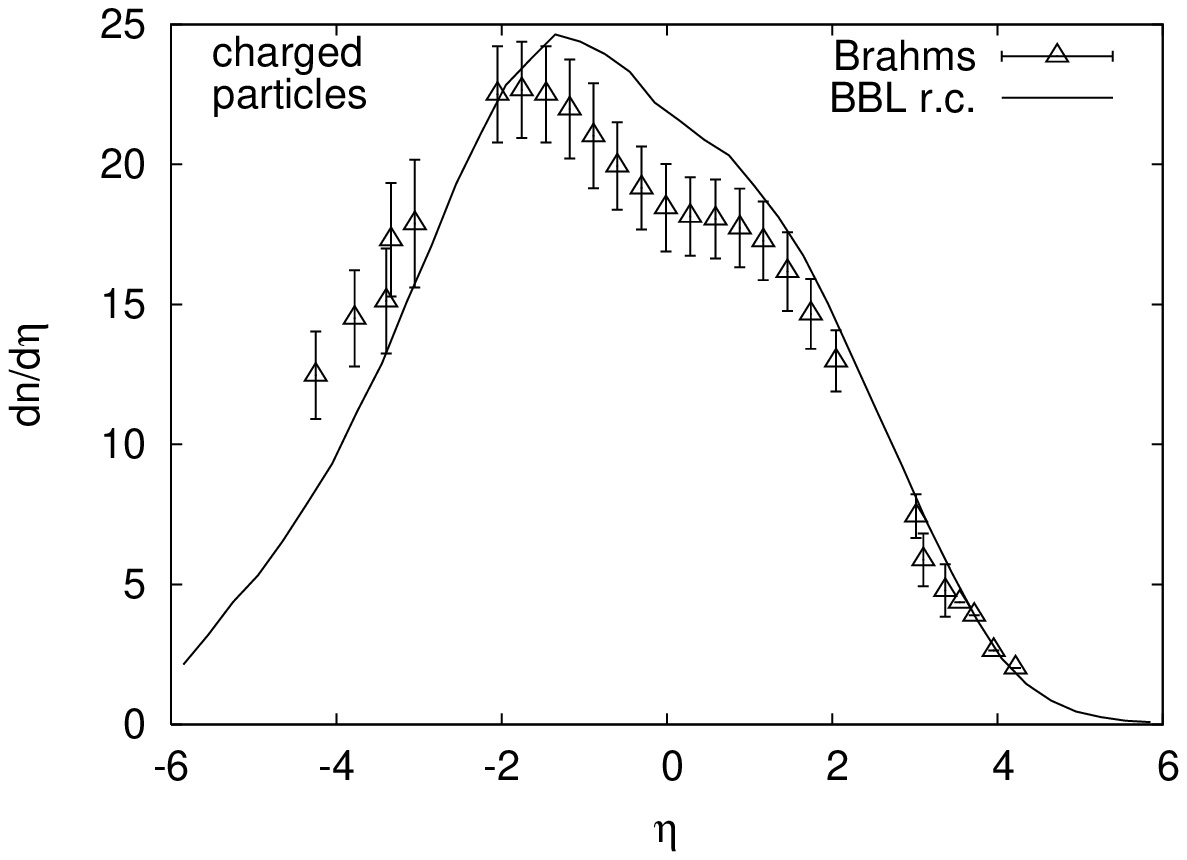}
\includegraphics[width=\ewidth]{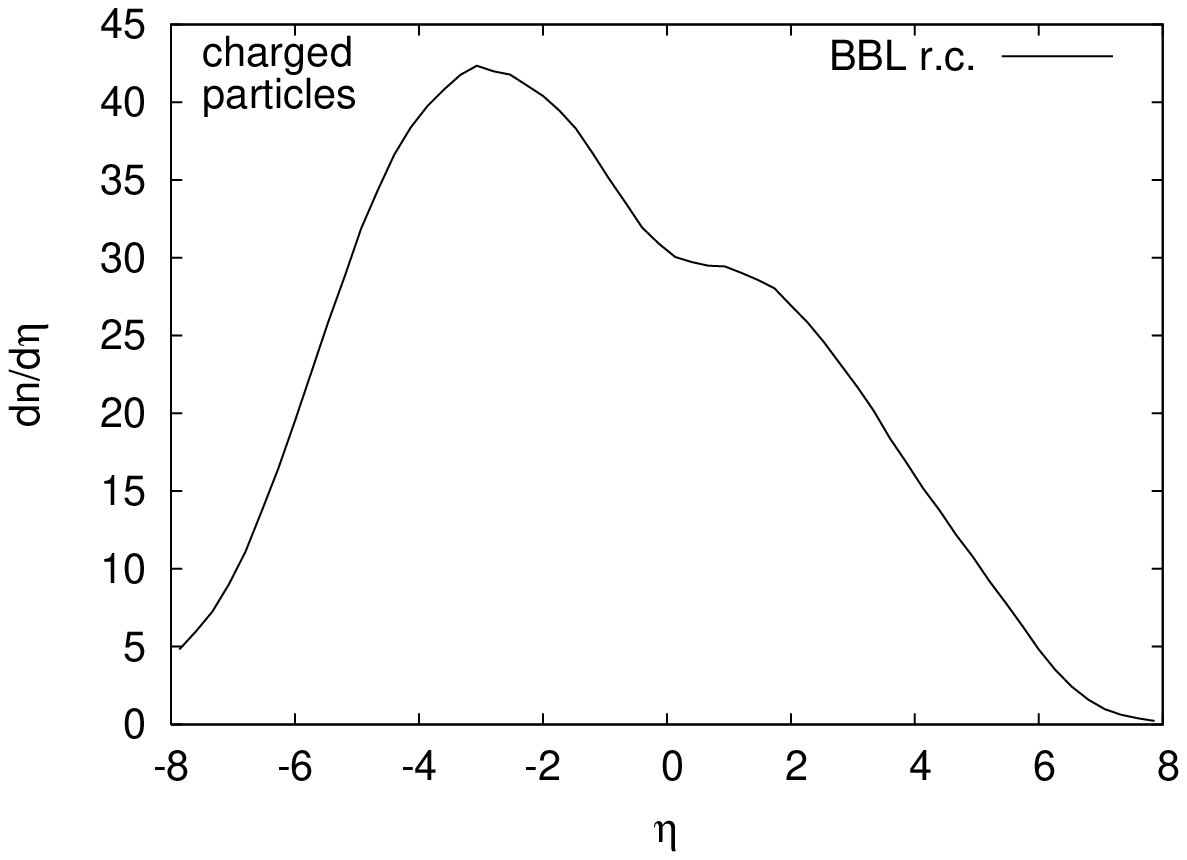}
\caption{Left: Comparison of the BBL event generator (with running
  coupling evolution) to RHIC data ($E_{\rm CM}=100$~GeV) from
  $d+Au$ collisions by the BRAHMS
  collaboration~\protect\cite{BRAHMS}. We have scaled our results for
  $p+Au$ (obtained with $N_A = 6$ participants) by a factor of two.
Right: our prediction for central $p+Pb$ collisions at LHC energy,
   (assuming $N_A=10$ and $E_{\rm CM}=3000$~GeV).
 \label{fig:pAu}}
\end{figure}
As a first check, we apply our model to RHIC energy which represents
the highest presently available energy for proton-nucleus collisions.
We compare the (pseudo-) rapidity distribution of inclusive charged
hadrons to data by BRAHMS\cite{BRAHMS}. The fragmentation region of
the target should be disregarded since no attempt has been made to
treat that realistically.  Given that no special tuning has been
performed to fit these particular data\footnote{For example regarding
diquark recombination, string fragmentation or initial conditions for
the evolution of the two saturation momenta}, we consider the
qualitative agreement to be quite good. More importantly, we note that
both evolution scenarios (running and fixed coupling) can easily be
made to fit the same data at this energy by somewhat readjusting the
initial conditions for $Q_s$ (see for example\cite{KL} for a much
better fit than ours with fixed-coupling evolution). Hence,
RHIC energy is too low to reliably probe the evolution of $Q_s$;
rather, results are mostly sensitive to the initial conditions. 

In the
right panel we show our result for central $p+Pb$ collisions at LHC
energy, which roughly agrees with that from ref.\cite{KL_LHC}. Note,
however, that ref.\cite{KL_LHC} considers the overall normalization to
be a parameter, fixed from low-energy (RHIC) collisions, while in
our approach it is determined automatically by momentum
conservation. The similarity of $dN/d\eta$ at central rapidities
obtained via the two methods perhaps
suggests that indeed the number of radiated gluons equals the
maximum number allowed by kinematics.

\begin{figure}[htb]
\begin{center}
\includegraphics[width=\ewidth]{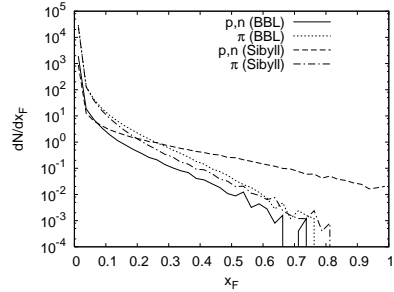}
\caption{Comparison of the \sibyll~and BBL models for central
  $p+^{14}\!N$ events at $E=10^9$~GeV. For either model, we show the
  nucleon and pion spectra separately. 
One observes the suppression of forward nucleon
  production in the high-density limit, which is due to
  the complete breakup of the proton. \label{fig:xf0}}
\end{center}
\end{figure}
Figure~\ref{fig:xf0} shows the $x_F$ distribution of pions and
nucleons for central proton-nitrogen events at $10^9$~GeV. We plot on
a logarithmic scale to show the effect at high $x_F$.  Very forward
particle production is suppressed as compared to the pQCD model
\sibyll, which is a consequence of the break-up of the projectile into
its partonic components. This behavior affects another key quantity
for cosmic ray air showers, the so-called inelasticity, which is one
minus the Feynman-$x$ of the most energetic secondary hadron (shown
below). Also note that in the high-density limit diquark recombination
in the forward region is suppressed (see discussion below),
and so the projectile proton
mainly decays into a beam of leading mesons\cite{DGS}. It would be
very useful to check this prediction in central $p+Pb$ collisions at
the LHC in order to confirm or rule out the basic mechanism of energy
degradation presented here,
which is very important for cosmic ray air showers.

\begin{figure}[htb]
\includegraphics[width=\ewidth]{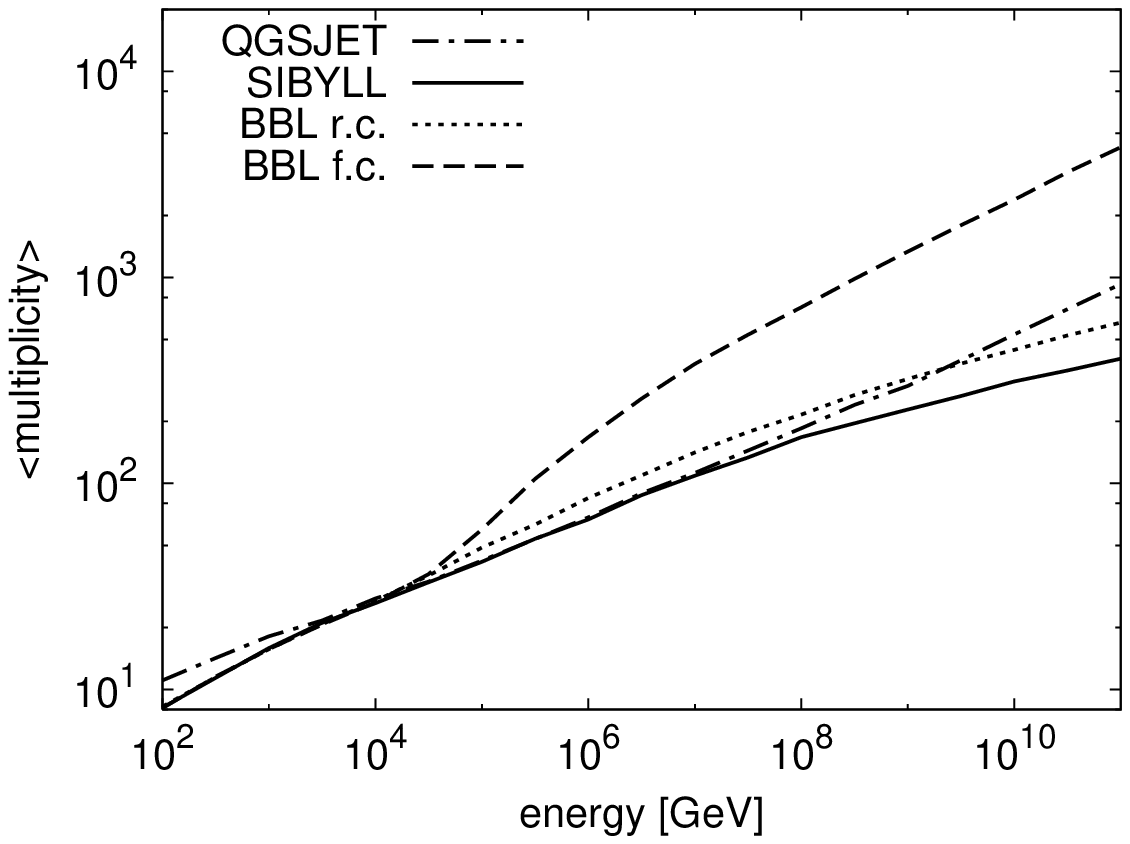}
\includegraphics[width=\ewidth]{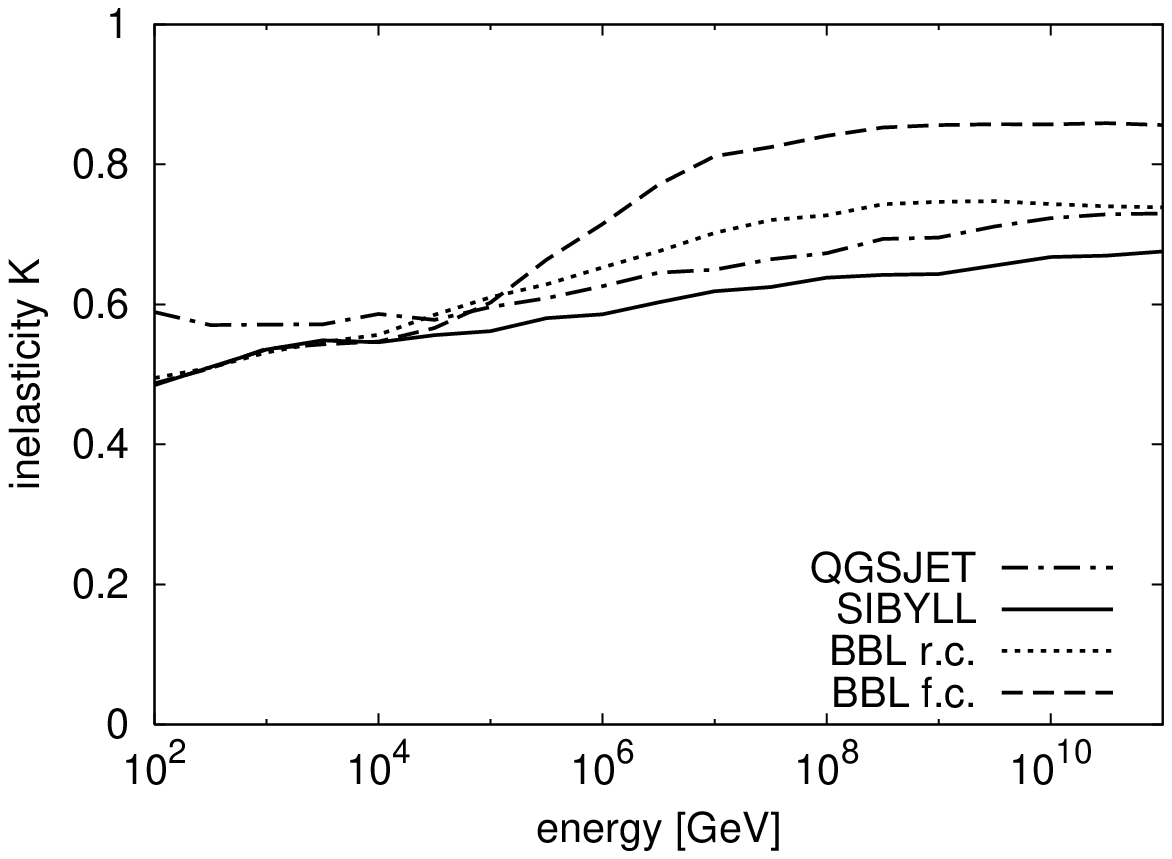}
\caption{Mean multiplicity of charged particles  (left panel) and the
  inelasticity (right panel) as a function of lab energy, for minimum
  bias $p+^{14}\!N$ collisions.\label{fig:mult}}
\end{figure}
Fig.~\ref{fig:mult} compares the mean multiplicities of charged
particles from the BBL model with fixed and running coupling evolution
of the saturation scale, and the conventional models \sibyll~and \qgsjet. Most
significant is the difference between fixed and running coupling
evolution. For fixed coupling evolution $Q_s^A$ grows very large over
a broad range of impact parameters (the radius of the black disc
approaches the geometrical cross section of the target nucleus).
Hence, even for minimum bias collisions forward particle production is
strongly suppressed in this case, and the remaining energy is used for particle
production at small $x_F$. This explains the large multiplicities as
compared to BBL with running coupling evolution. On the other hand,
for energies below $10^4-10^5$~GeV there is little sensitivity to the
evolution scenario for $Q_s^A$ and the results look rather similar.
In {\sibyll,} the growth
of the multiplicity is ``tamed'' by a rapidly growing $p_t$-cutoff for
leading-twist hard processes. It should be kept in mind though that
the multiplicity from large momentum transfer processes is power-law
sensitive to the infrared cutoff.

\begin{figure}[htb]
\includegraphics[width=\ewidth]{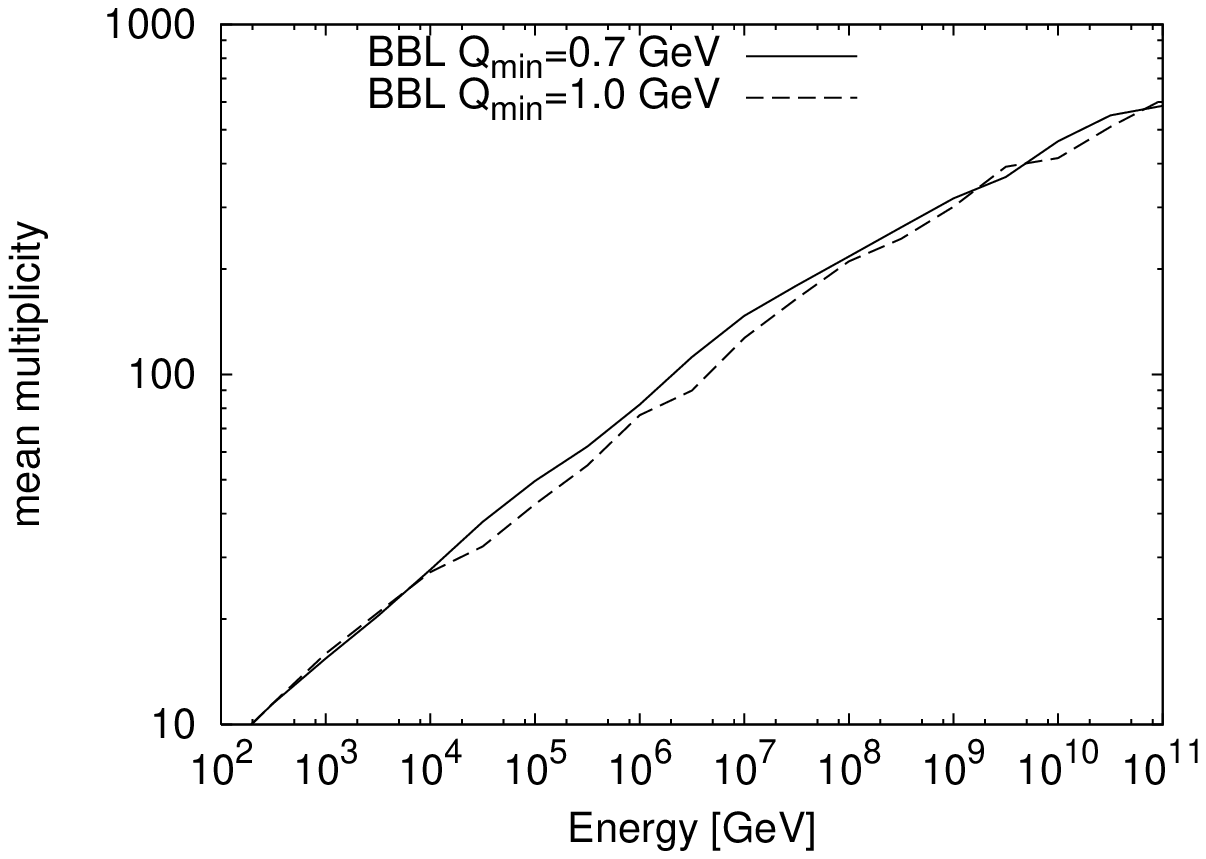}
\includegraphics[width=\ewidth]{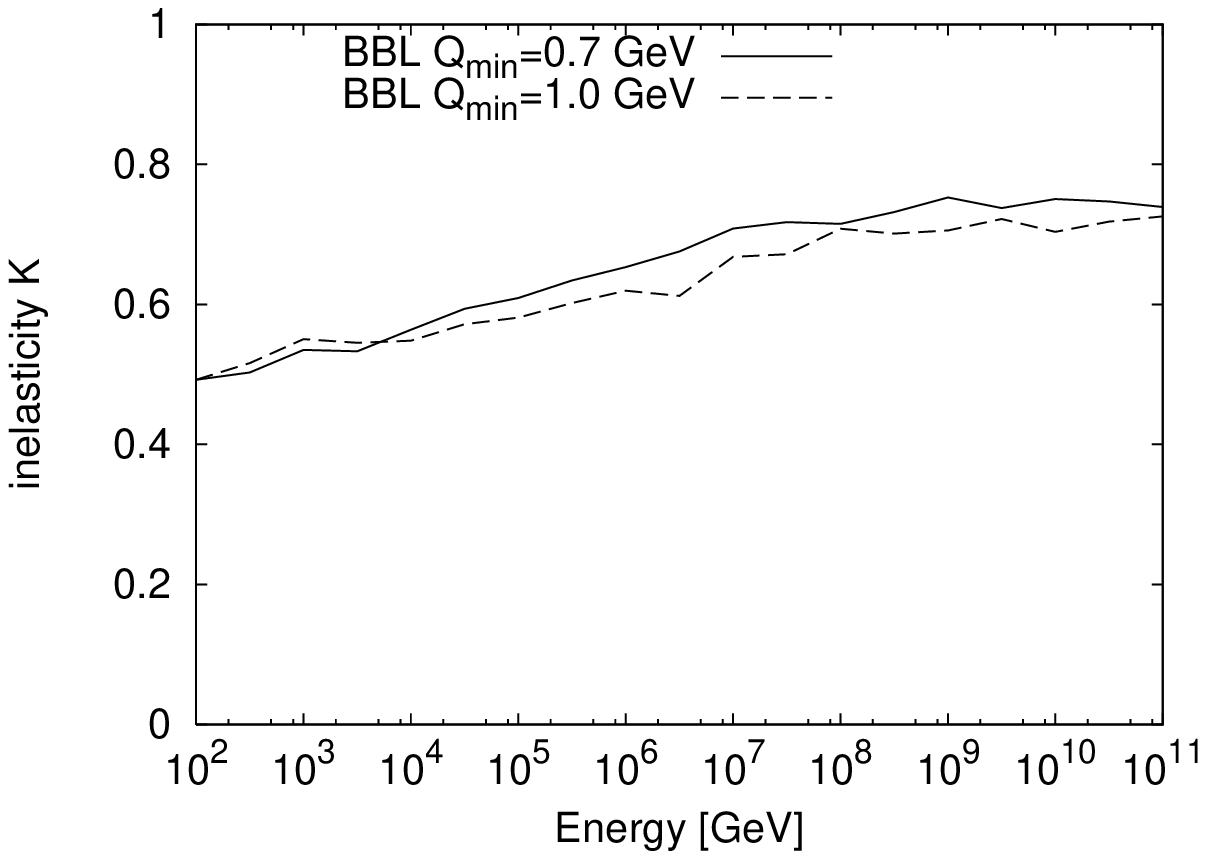}
\caption{The mean charged particle multiplicity and the inelasticity
  for minimum-bias $p+^{14}\!N$ collisions using the combined
  BBL+\sibyll~ model for
  $Q_{\min}=0.7,~1.0$~GeV. \label{fig:mult1}}
\end{figure}
To check the sensitivity to the (artificial) boundary between low and
high density from eq.~(\ref{Qs_xmin}), we compare results for the
multiplicity and for the inelasticity for $Q_{\min}=0.7$~GeV and
$Q_{\min}=1$~GeV in Fig.~\ref{fig:mult1}.  A lower value for
$Q_{\min}$ leads to a higher fraction of BBL events, but these are
then generated with lower $Q_s^A$, and so are more similar to ``soft''
events from SIBYLL. In total, we see that there is little sensitivity
of physical observables to the precise threshold between the models,
as long as it is chosen within reasonable bounds. In the following we
chose $Q_{\min}=0.7$~GeV as default for our calculations.

\begin{figure}[tb]
\includegraphics[width=\columnwidth]{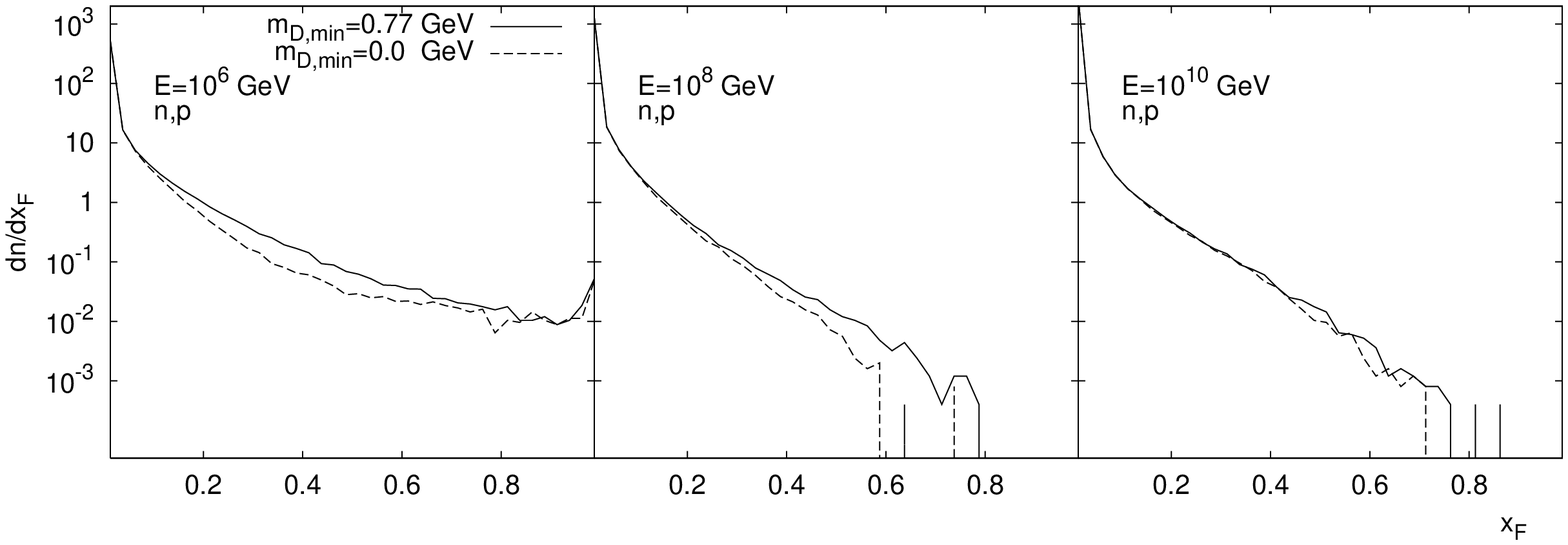}
\caption{Diquark recombination effect for $p+^{14}\!N$ collisions at various
  energies.\label{fig:diquark}} 
\end{figure}
In Fig.~\ref{fig:diquark}, finally, we show the effect of the diquark
recombination mechanism.  We compare the production of protons and
neutrons in central collisions to the case without recombination
($m_{\rm cut}=0$). At relatively low energies ($E\approx 10^{6}$~GeV),
one notices a suppression of forward baryon production when
recombination is not taken into account, except for $x_F\approx
1$. This very forward peak is in fact produced by elastic or
diffractive events within the SIBYLL model, which still handles
about 5\% of all central collisions at this energy if $Q_s$ obeys running
coupling evolution\footnote{We note that the actual configuration of
nucleons in the target is generated randomly in each event by SIBYLL
according to the appropriate nuclear density profile. Hence,
fluctuations in the number of target participants are taken into
account.}.  At $E=10^8$~GeV and above, essentially all central
collisions occur near the black body limit, i.e.\ they pass
criterion~(\ref{Qs_xmin}).  The diquark recombination mechanism then
allows more particles to be produced in the forward region since the
momenta of the corresponding valence quarks are combined. Nevertheless, the
effect is less important at higher energies since there $Q_s^A$ is
already too high to allow for the 
production of a diquark system with low invariant mass.


\section{Air showers} \label{sec_AirSh}
In section~\ref{sec_IntroAirSh} we give a brief introduction into
concepts and observables in cosmic ray physics for readers from other fields.
More detailed discussions can be found
in dedicated textbooks such as the book by Gaisser\cite{GaisserBook}.

In section~\ref{AirShSim} we discuss general aspects of air shower
simulation and, in particular, present the so-called cascade
equations\cite{seneca} employed here to solve for the longitudinal shower
profile. From those equations, we can cleanly identify which ``input''
is required from QCD for their solution and, indeed, which properties of
high-energy hadronic interactions actually influence the characteristics
of very high energy air showers.

\subsection{Introduction to air showers} \label{sec_IntroAirSh}
 
Due to a very low flux at high energies, cosmic rays are detected
indirectly by the measurement of air showers. These are cascades of
particles produced by the interaction of the primary cosmic ray and
subsequent secondaries with air nuclei. An air shower can be
structured into 3 parts: a hadronic, an electromagnetic and a muonic
part. Hadrons are produced in collisions with air-nuclei.  Most of
the electromagnetic part is induced by $\pi^0$-decays, which have a
short life-time and decay instantly up to $10^{19}$~eV. Muons are
produced by decays of charged pions and kaons, but since their
decay-length is much longer only low energies particles decay while at
higher energies collision with air nuclei dominates. Once produced,
muons propagate with little interaction (mostly energy-loss) through
the atmosphere until they decay or reach the ground.
The most prominent fraction of a shower is the electromagnetic part. 
A rule of thumb is that the number of electrons and positrons
at the maximum of the shower is approximately 60\% of the primary energy
$E_0$ measured in GeV (a $10^{11}$~GeV shower produces about 60 billion
particles~!).

\begin{figure}[t]
\includegraphics[width=\ewidth]{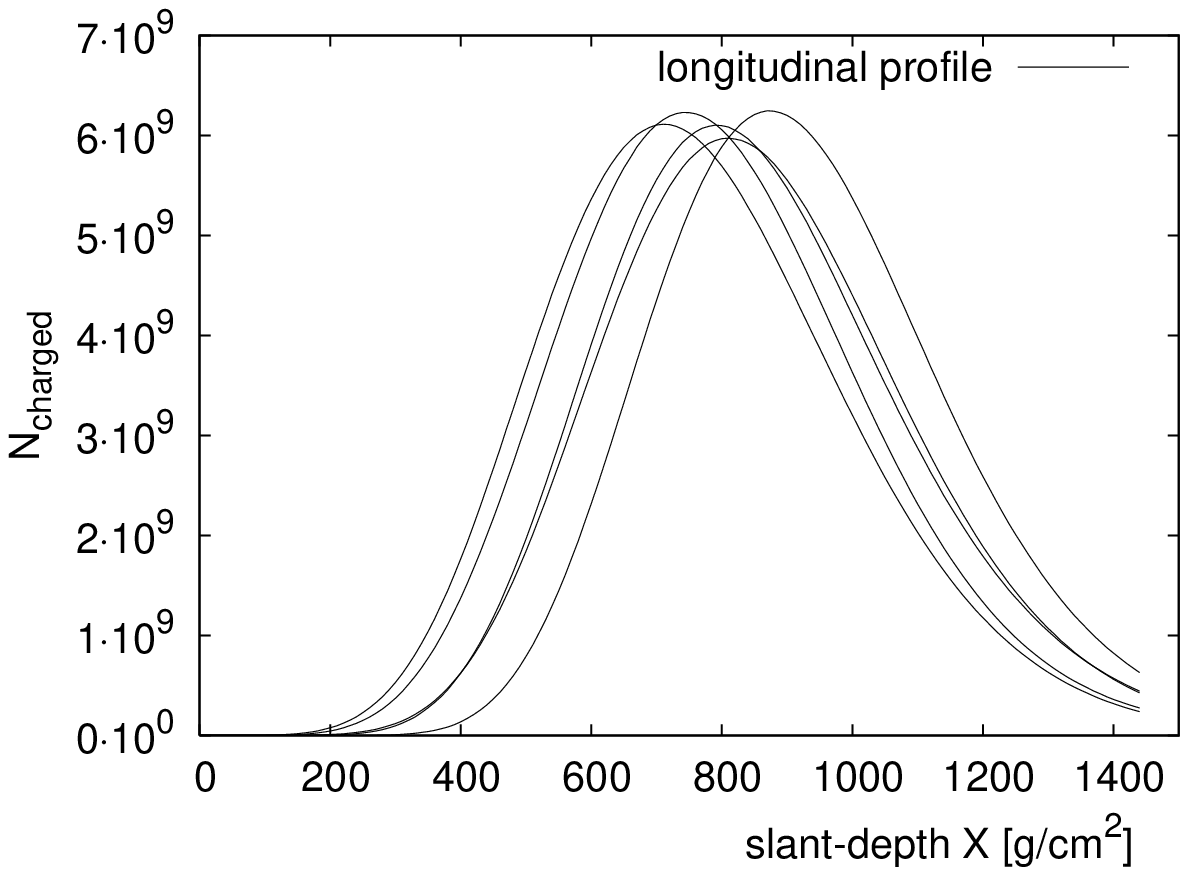}
\includegraphics[width=\ewidth]{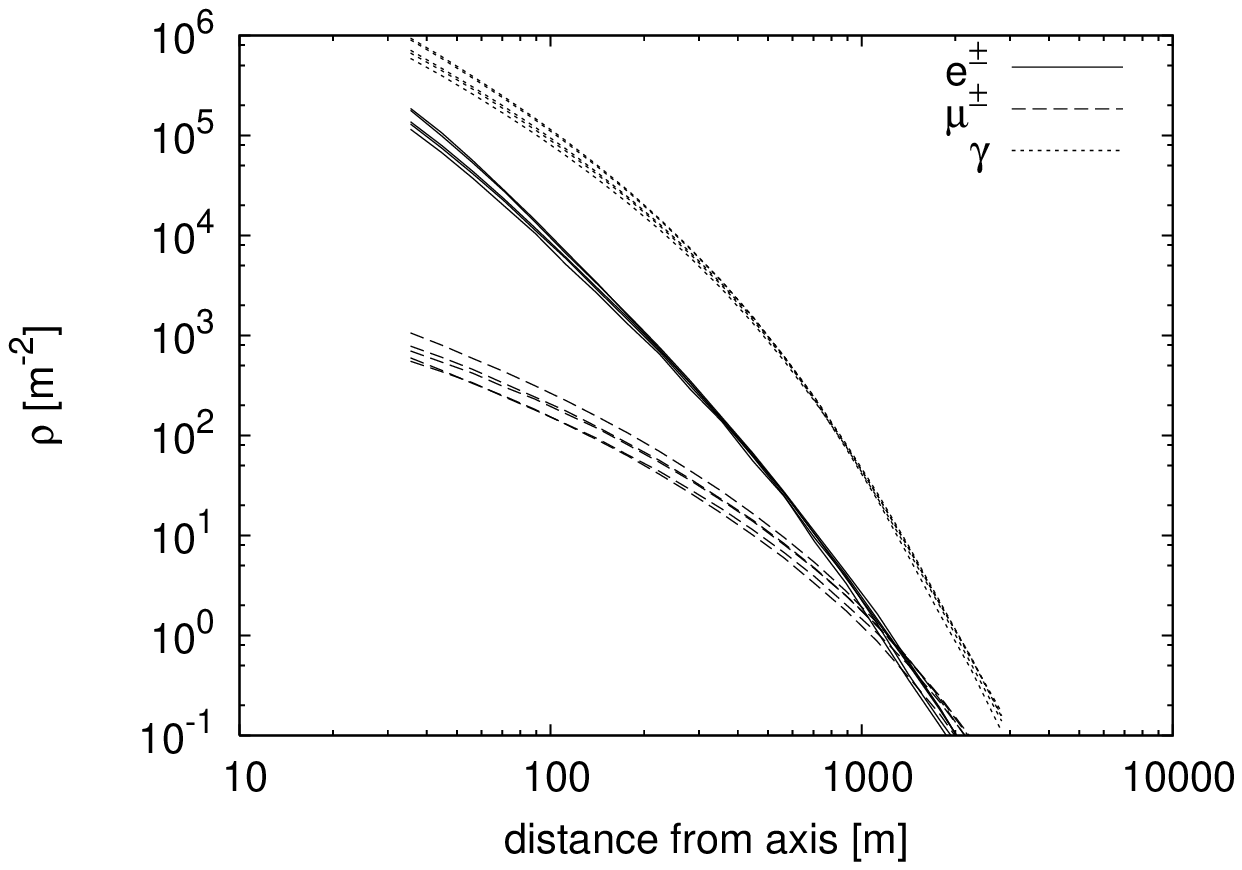}
\caption{Longitudinal profiles and lateral distribution function of 
  typical showers; here $E_0=10^{10}~GeV$.\label{fig:profile}} 
\end{figure}
There are two basic observables associated with air showers, which
are measured by experiments: the longitudinal shower profile and the
lateral distribution functions. The longitudinal profile is the number
of charged particles measured along the shower-axis. 
One typically expresses this as a function of slant-depth, which is
the density of the atmosphere integrated along the
shower-axis: $X=\int_\infty^{p} \rho_{\rm Air}(l)dl$. For a vertical
shower, $X$ ranges from zero (top of the atmosphere) to 1020~g/cm$^2$
(sea level). For inclined showers with polar angle
up to 60$^\circ$, the
slant depth is related to the vertical depth by $X=X_v /
\cos(\theta)$; at larger angles one has to take into account the curvature
of the earth. Typical shower profiles are shown in
Fig.~\ref{fig:profile}. The position where the profile reaches the
maximum is defined as the shower maximum $\Xmax$, the number of
charged particles is called the shower size $\Nmax$. For a fixed
energy these values fluctuate quite substantially, which is
 due to the fact that the depth of
the first collision can vary according to the cross section. One
therefore usually compares the mean $\Xmax$ for a given primary and energy
$E_0$.
An important observation is that $\Xmax \propto \log(E_0)$ and $\Nmax
\propto E_0$. To 
first approximation,  the shower of a nucleus can be considered
to be a superposition of $A$ independent nucleon-initiated showers,
each carrying an energy
$E_0/A$ ($E_0$ is the total energy of the primary, not per
nucleon). The mean $\Xmax^{\rm Fe}$ of an iron-induced shower is
therefore lower than that of a proton induced shower, roughly
corresponding to the mean $\Xmax$ of a proton shower at energy $E_0/A$. 
Experiments measure the longitudinal profile via the emitted
fluorescence light of nitrogen as the shower swipes through the
atmosphere. The energy of the primary cosmic ray is then proportional
to the total number of charged particles, which is determined by
integrating the profile. 

A very important quantity for $\Xmax$ is the inelastic
cross section of a particle on air. It
determines the mean free path in the atmosphere. A significant
amount of uncertainty in models for the longitudinal distribution
of air showers is due to this variable.  

The other observable, the lateral distribution function (LDF),
describes the density of particles measured on the ground as a
function of the distance from the shower axis (hence in the
shower plane) for given particle types\footnote{Experiments measure
the density in terms of response of whatever detector they use, e.g.\
scintillation or Cerenkov light, and normalize by the average signal
of atmospheric muons, which are used for calibration.}. Most of the
lateral spread is generated by low-energy scattering of the
electromagnetic part of an air shower. Hadrons do not spread out very
much to large distances, only the low energy ones influence the tail
of the LDF by producing $\pi^0$ at large angles or
distances\footnote{Hence, LDFs constrain mostly the {\em low-energy}
  hadronic interaction models\cite{hadrLDF}.}.  Typical
LDFs are depicted in Fig.~\ref{fig:profile} (right panel). They follow
approximately a power law.  Just as $\Xmax$, the slope of the LDF also
fluctuates. Showers induced higher in the atmosphere lead to flatter
LDFs since they spread out over a larger radial distance.
Empirically, one finds that these fluctuations cancel at some distance
from the shower axis. Experiments exploit this property to extract the
primary energy of the cosmic ray, which is taken to be proportional to
the density at some distance from the axis. The proportionality constant
is normally computed from simulations.

When studying high energy particle physics with air showers it is
important to notice that the high-energy part of the shower (i.e.\ the
first collisions) is almost
purely longitudinal, which follows from simple kinematics. The
targets (air-nuclei) are at rest and the projectiles have huge
$\gamma$-factors, resulting in very small scattering
angles. Furthermore, 
forward scattering is most important in air showers, since large-$x_F$
particles carry most of the energy. This implies, for example, that
high-$p_t$ QCD jets at mid-rapidity do not influence the longitudinal
shower profile substantially nor do they contribute significantly to
the lateral spread (see also ref.\cite{GaisserHalzen}).

\subsection{Simulation of air showers} \label{AirShSim}

The simulation of air showers is crucial for cosmic ray physics, since
they are needed for the interpretation of experimental data. 
Given a hadronic interaction model\footnote{Typically a Monte-Carlo
  event generator which generates complete final states and accounts
  for fluctuations.}, and 
models for electromagnetic and muonic interactions
one could just follow each particle and subsequent secondaries
individually. Of course, since $N\propto E_0$, this would require huge
amounts of computing time at very high energies. 
Therefore, Hillas introduced the thinning algorithm:
below a given energy threshold, i.e.\ for $E<f_{\rm th} \times E_0$, only one
single secondary particle from a collision is followed, but it is attributed a
higher weight. However, for a large thinning level $f_{\rm th}$, this
introduces artificial fluctuations into the air shower. 

On the other hand, the fact that at high energies the lateral part of the
hadronic shower can be neglected suggests another efficient approach to
solve this problem, which is based on one-dimensional transport
equations\cite{seneca}:
\begin{eqnarray}
\frac{\partial h_{n}(E,X)}{\partial X} &=&  -h_{n}(E,X)\left[
  \frac{1}{\lambda _{n}(E)}+\frac{d_{n}}{E\rho(X)}\right] \label{for:hce} \\ 
  +\sum _{m} & & \int_{E}^{E_{\mathrm{max}}} h_{m}(E',X)
 \left[
  \frac{W_{mn}(E',E)}{\lambda _{m}(E')} \right. \nonumber 
 \left. + \frac{d_{m}D_{mn}(E',E)}{E'\rho(X)}\right] dE'\nonumber ~~. 
\end{eqnarray}
Here,
$h_n(E,X)dE$  is the number of particles of type $n$ at altitude $X$
in the given energy range $[E,E+dE]$;
the functions \( W_{mn}(E',E) \) are the energy-spectra \(
{dN}/{dE} \) of secondary particles of type \( n \) in a collision
of hadron \( m \) with air; \( D_{mn}(E',E) \) are the
corresponding decay functions; $d_n=m_n/(c\tau_n )$ is the decay
constant, and $\lambda_{n}(E)\propto 1/\sigma_{\rm inel}$ is the mean
free path of the particle.
The first term in~(\ref{for:hce}) with the minus sign accounts for
particles disappearing by either collisions or decays, whereas the source
term accounts for production of secondary particles by collisions or
decays of particles at higher energies.
Primary particles appearing in eq.~(\ref{for:hce}) are nucleons
(protons, neutrons and their anti-particles), charged pions, charged
and neutral kaons. In addition, we have as secondaries 
$\pi^0$s and photons (as direct decay products from $\eta$ mesons, for
example) 
which feed the electromagnetic cascade and muons as decay products
of charged mesons. 

To summarize, the basic ingredients for constructing longitudinal
profiles of air showers are the inclusive spectra $dN_n/dx_F$ of
the non-strongly decaying particles and their inelastic cross sections,
which determine the mean free path. The electromagnetic cascade can be
treated in a similar way\cite{EGS4}.

The first few interactions in an air shower are the main source of
fluctuations in $\Xmax$ (and, accordingly, in the LDF). Since the
cascade equations cannot account for those (they solve for a mean
shower) one could treat the high energy part by a traditional
Monte-Carlo method. This is the so-called hybrid approach to air
shower simulations. On the other hand, if one solves the cascade
equations without fluctuations\cite{pylos} one can still reproduce the
average $\Xmax$ to within a few g/cm$^2$.

\subsection{Sensitivity of $\Xmax$ to the $x_F$ distribution} 
\label{sec_XmaxSensitiv}

Finally, we analyze which region of the
$x_F$-distribution is most important for the mean $\Xmax$ of an air
shower. From the simple argument that forward particles carry
most of the energy it should be clear that the high $x_{F}$ region is
important. Our goal here is to quantify this statement somewhat. 

Given $dN_n/dx_F$ distributions of secondaries, to study the
sensitivity of $\Xmax$ to various regions of $x_F$ we solve the
cascade equations with a modified distribution: 
\be  \label{modif_dNdxF}
\frac{dN_n}{dx_F}
\rightarrow \frac{dN_n}{dx_F} (1+\epsilon) {\rm ~~for~} x_F<x_F^0~.  
\ee
That is, we enhance or suppress the spectra at $x_F<x_F^0$
relative to the default reference distributions, depending on the sign
of $\epsilon$. At the same time, we suppress or enhance 
particles at $x_F>x_F^0$
in such a way as to conserve the total energy: 
\be
\frac{dN_n}{dx_F}
\rightarrow \frac{dN_n}{dx_F} (1-\epsilon') {\rm ~~for~} x_F>x_F^0~,
\ee
with
\be
\epsilon' = \epsilon~\frac{\sum\limits_n\int\limits_0^{x_F^0} dx_F 
   ~E_n(x_F)~\frac{dN_n}{dx_F}}
{\sum\limits_n\int\limits_{x_F^0}^1 dx_F ~E_n(x_F)~\frac{dN_n}{dx_F}}~.
\ee
Note that the $dN_n/dx_F$ are, of course, energy dependent while we take the
$1+\epsilon$ factor in~(\ref{modif_dNdxF}) to be constant. Also, we do
not modify the inelastic cross section (i.e.\ the mean free path in
the atmosphere), just the $x_F$-distribution of secondaries in an
inelastic event.
We then solve eqs.~(\ref{for:hce}) to determine the change of $\Xmax$
relative to that for the reference distributions as a function of both
$\epsilon$ and $x_F^0$. 

\begin{figure}[htp]
\begin{center}
\includegraphics[width=\ewidth]{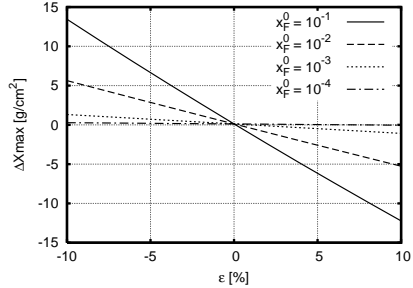}
\end{center}
\caption{The shift of $\Xmax$ as a function of $\epsilon$ for various
  $x_F^0$. This shows that the shower maximum is sensitive to mainly
  the forward region, $x_F\gsim10^{-3}$. \label{fig:sens}}
\end{figure}
The result is shown in Fig.~\ref{fig:sens}, assuming a proton primary
with energy $E_0=10^{10}$~GeV.  The reference $dN_n/dx_F$ distributions
were taken from \qgsjet. 
We observe that $\Delta\Xmax$ is approximately linear in $\epsilon$.
For large $x_F^0$, for example =0.1, there is a significant shift
$\Delta\Xmax\approx 140 \epsilon$. For $\epsilon<0$ we suppress the
$x_F<0.1$ region and, by energy conservation, enhance the large-$x_F$
part; this leads to deeper penetration into the atmosphere, i.e.\ to
larger $\Xmax$. In turn, {\em suppression} of forward particle
production ($\epsilon>0$ and $\epsilon'<0$) leads to decreasing
$\Xmax$.

However, one can also observe that
$\Xmax$ becomes independent of $\epsilon$ for $x_F^0\lton
10^{-3}$. This shows that the small-$x_F$ part of the distribution has
no influence on the shower maximum (for fixed cross section).
For comparison, we note that a
particle produced at mid-rapidity (in a collision with energy
$10^{10}$~GeV) with an energy of about the proton mass has $x_F\simeq
10^{-5}$.


\section{Application of the BBL to air showers}

\begin{figure}[htp]
\includegraphics[width=\ewidth]{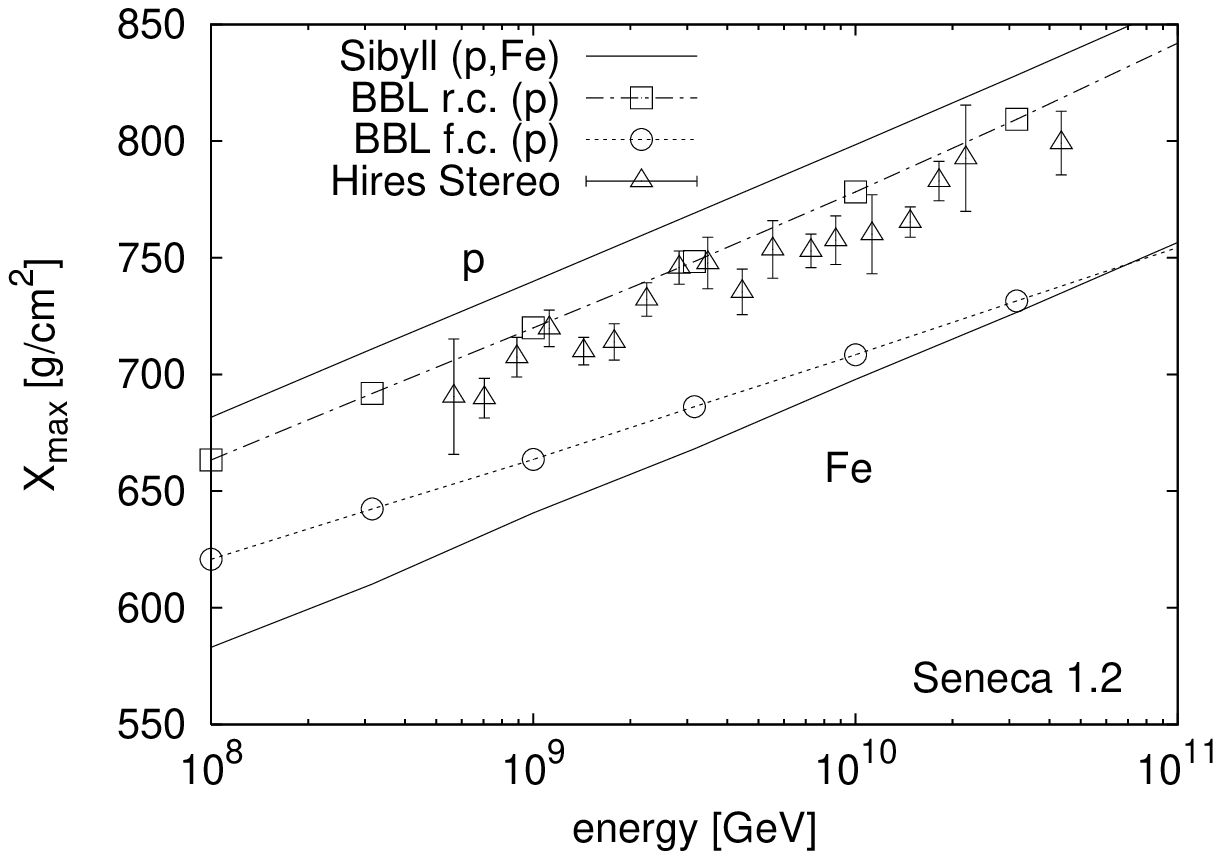}
\includegraphics[width=\ewidth]{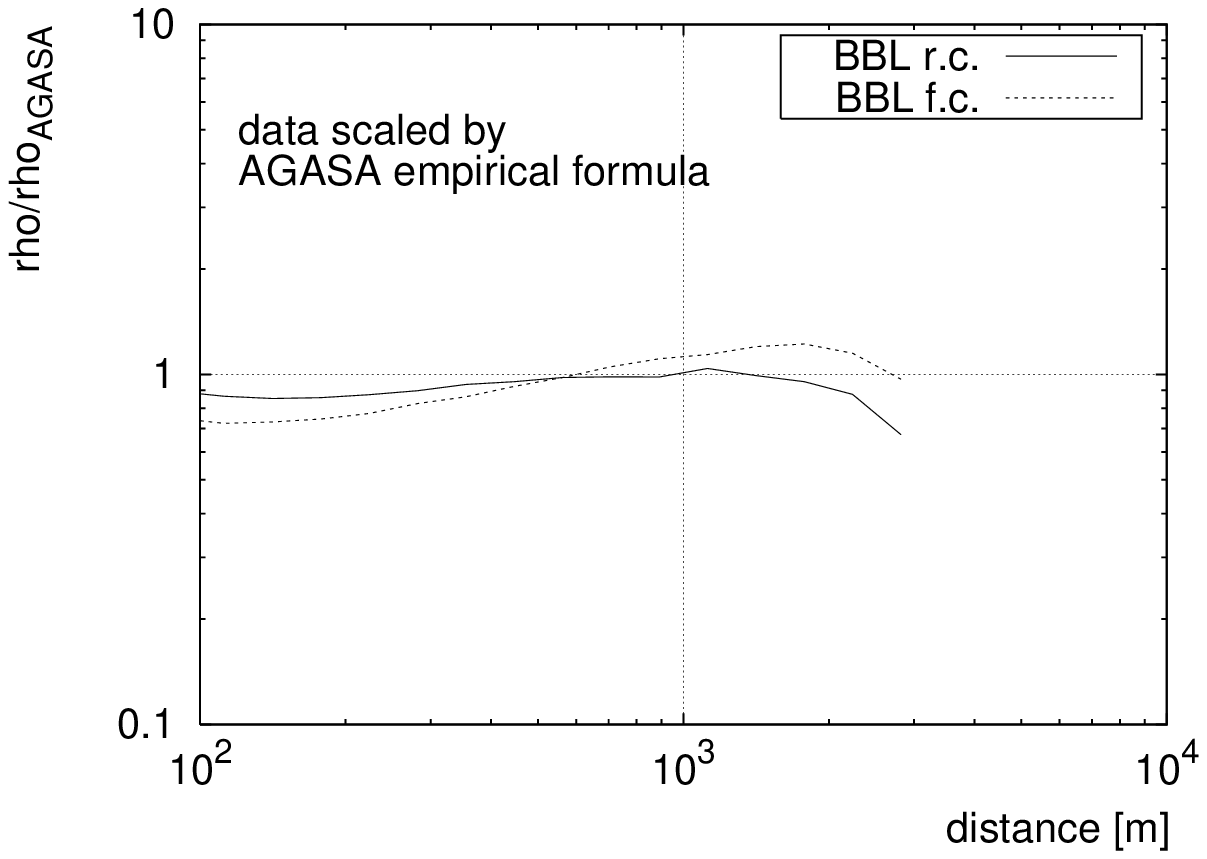}
\caption{Left panel: mean $\Xmax(E_0)$ for $p$ and $Fe$ induced
showers from {\sibyll,} and for $p$ primaries with BBL for fixed and
running coupling evolution; data are from HIRES\protect\cite{hires};
figure from\protect\cite{DDS}.
 Right panel: mean lateral distribution function scaled by the
 LDF found by AGASA.
 \label{fig:xmax}}
\end{figure}

To apply our interaction model to air showers,
we tabulated $dN_n/dE$ distributions of the
primary particles appearing in an air shower. We then employ the
\seneca\cite{seneca} model to solve the cascade equations
(\ref{for:hce}). The hadron-air inelastic cross sections are taken as
parameterized in \sibyll.
In Fig.~\ref{fig:xmax} we compare the results for
fixed and running coupling evolution to those obtained with the
\sibyll~model.
One notices a huge difference between fixed and running coupling
evolution scenarios. The saturation momentum in the former case is so
high that forward scattering is very strongly suppressed over a broad
range of impact parameters; also, the discrepancy
between those evolution scenarios at the highest energies is strongly
amplified by subsequent hadronic collisions in the cosmic ray
cascade. Consequently, for fixed coupling evolution the shower is
absorbed very early in the atmosphere. Hence, 
if we assume a hadronic primary, then the HIRES\cite{hires} data excludes
this scenario, since it would require hadrons lighter than protons. 
This is a novel result as present accelerator data could not
rule out such a rapid growth of the gluon density (see
e.g.\cite{gbw,KL}); it illustrates the ability of cosmic ray air showers
to provide observational constraints on small-$x$ QCD~\cite{DDS}.

The running coupling result, on the other hand, is compatible with
those data and with a light composition. The results from this
model are similar to those from \sibyll~or \qgsjet~(for proton
primaries) within present theoretical uncertainties. Nevertheless, our
results show that the effects discussed here make near-GZK proton-induced air
showers look more similar to nucleus-induced cascades from leading-twist
models and so favor a lighter composition near the cutoff.

Lateral distribution functions obtained with both evolution scenarios 
are shown in the
right panel of Fig.~\ref{fig:xmax}. The LDFs were computed for the
AGASA experiment and include a full detector response simulation of
the plastic scintillators. The results are scaled by the empirical
formula which describes the data up to highest energies quite well:
\begin{equation}
S(R)=C\left(\frac{R}{R_{M}}\right)^{-\alpha}\left(1+\frac{R}{R_{M}}
\right)^{-(\eta-\alpha)}\left(1+\left(\frac{R}{1~{\rm 
km}}\right)^{2}\right)^{-\delta}, \label{for:LDF}
\end{equation}
with $\alpha=1.2$, $\delta=0.6$,  $R_{M}=91.6$~m and $\eta=3.84$ for
vertical showers\cite{Takeda:2002at}.
The parameter $C$ is adjusted using the energy conversion formula~(13)
from Ref.~\cite{Takeda:2002at}, 
\begin{equation}
E=2.17\times10^{17}~S(600~{\rm m})\,\mathrm{eV}\, \label{for:engy},
\end{equation}
which is valid for the average altitude of the AGASA array, =667~m.  The
comparison in Fig.~\ref{fig:xmax} shows that the LDF obtained for
fixed coupling evolution is much flatter than that for running
coupling evolution, which in turn agrees better
with the data (notice that in the figure, the theoretical curves are
scaled by the data).
 This is consistent with our finding for $\Xmax$, as
discussed above. When the
shower is absorbed earlier in the atmosphere then
it  spreads out to larger radial distances from the shower axis.


\section{Conclusion and Outlook}

In this paper we pointed out that atmospheric air showers induced by the
highest energy cosmic rays are sensitive to QCD interactions at
extremely small momentum fractions $x$ where nonlinear effects are
expected to play a major role and lead to unitarization of partonic
scattering cross sections. In turn, this means that cosmic rays air
showers can provide valuable insight and observational constraints
for the strong-field regime of QCD. As an example, we have shown that
present data on $\Xmax(E)$ already exclude that the rapid $\sim
1/x^{0.3}$ growth of the saturation boundary (which is compatible with
RHIC and HERA data) persists up to GZK cutoff energies\cite{DDS}.

The model used here for quantitative calculations can be improved in
many ways, for example by incorporating more advanced estimates for
the small-$x$ gluon densities obtained from the approaches of
refs\cite{Altarelli:2003hk,{Ciafaloni:2003rd}}. From the point of view
of learning about cosmic rays from small-$x$ QCD it could be
interesting to extend the studies to nucleus-nucleus collisions and to
perform a composition analysis near the cutoff. This might also
be relevant for physics well below the cutoff, in the region above the
``knee''\cite{Kampert:2004rz} ($E\simeq 5\cdot10^6$~GeV) because for nuclei
nonlinearities should set in at lower energies already.

Cosmic ray air showers offer several important advantages over laboratory
experiments: first of all, of course, their energies can exceed those
of accelerators (even LHC) by far. Second, many properties of extensive
air showers are sensitive mainly to the forward region and to
transverse momenta about $\langle p_t\rangle$, which means that they
probe extremely small $x$ in the target nucleus. Finally, an air
shower develops via several subsequent collisions and so any
``distortion'' of the momentum-space distribution of secondaries from
high-density effects is strongly amplified (essentially raised to the
power of the number of collisions). 

On the other hand, unlike air shower detectors accelerator experiments
can control key parameters of the interaction. For example, aside from
collision energy and centrality one can also chose
various projectiles and targets, from protons
over light nuclei up to very massive nuclei such as gold or
lead. Central collisions on lead at LHC energy should
provide similar gluon densities as those on air at cutoff
energies. Hence, fruitful lessons regarding small-$x$ QCD will
hopefully emerge from both cosmic ray and accelerator data in the
future. We emphasize that crucial data to be obtained at the LHC is
not limited to the total proton-proton cross
section but includes $x_F$ distributions of secondaries from $p+A$
collisions in both the central and forward regions. The latter would
allow us to study the energy degradation mechanism in central
collisions, which plays an important role for cosmic ray air showers.

\section*{Acknowledgements}
This work was presented at the 44th INFN Workshop, ``QCD at Cosmic
Energies'', Aug.~29 -- Sept.~5,  2004, Erice, Italy;
http://www.lpthe.jussieu.fr/erice.
We thank Yu.~Dokshitzer, R.~Engel, L.~Frankfurt, T.~Gaisser,
J.~Jalilian-Marian and S.~Ostapchenko for stimulating discussions during
that workshop.

H.-J.D.\ acknowledges support by the German Minister for
Education and Research (BMBF) under project DESY 05CT2RFA/7.
The computations were performed at the
 Frankfurt Center for Scientific Computing (CSC).
The research of  M.S.\ was supported by DOE.


\begin{thebibliography}{0}

\bibitem{agasa}
S.~Yoshida  et al., Astropart.\ Phys.\ {\bf 3}, 105 (1995).

\bibitem {hires}
P.~V.~Sokolsky for the HiRes Collaboration,
{\it Prepared for 28th International Cosmic Ray Conference (ICRC 03), 
Tsukuba, Japan, July 31 -- Aug.\ 7, 2003}, p.\ 405;\\
High Resolution Fly's Eye Collaboration, arXiv:astro-ph/0407622.

\bibitem{auger}
Auger: www.auger.org

\bibitem{Altarelli:2003hk}
G.~Altarelli, R.~D.~Ball and S.~Forte, Nucl.\ Phys.\ B {\bf 674}, 459 (2003).

\bibitem{Ciafaloni:2003rd}
M.~Ciafaloni, D.~Colferai, G.~P.~Salam and A.~M.~Stasto,
Phys.\ Rev.\ D {\bf 68}, 114003 (2003).
 
\bibitem{bbl_gammaA}
L.~Frankfurt, V.~Guzey, M.~McDermott and M.~Strikman,
Phys.\ Rev.\ Lett.\  {\bf 87}, 192301 (2001).

\bibitem{DGS}
A.~Dumitru, L.~Gerland and M.~Strikman,
Phys.\ Rev.\ Lett.\  {\bf 90}, 092301 (2003)
[Erratum-ibid.\  {\bf 91}, 259901 (2003)].

\bibitem{DDS}
H.~J.~Drescher, A.~Dumitru and M.~Strikman, arXiv:hep-ph/0408073.

\bibitem{mq}
L.~V.~Gribov, E.~M.~Levin and M.~G.~Ryskin,
Phys.\ Rept.\  {\bf 100}, 1 (1983);\\
A.~H.~Mueller and J.~Qiu, Nucl.\ Phys.\ B {\bf 268}, 427 (1986);\\
I.~Balitsky, Nucl.\ Phys.\ B {\bf 463}, 99 (1996);\\
J.~Jalilian-Marian, A.~Kovner, A.~Leonidov and H.~Weigert,
Phys.\ Rev.\ D {\bf 59}, 034007 (1999)
[Erratum-ibid.\ D {\bf 59}, 099903 (1999)];\\
E.~Iancu, A.~Leonidov and L.~D.~McLerran,
Nucl.\ Phys.\ A {\bf 692}, 583 (2001);
A.~Kovner and U.~A.~Wiedemann,
Phys.\ Rev.\ D {\bf 66}, 034031 (2002).

\bibitem{sat}
L.~McLerran and R.~Venugopalan, Phys.\ Rev.\ D {\bf 49}, 2233 (1994);
ibid.\ {\bf 49}, 3352 (1994);\\
Y.~V.~Kovchegov, ibid.\ {\bf 54}, 5463 (1996);
ibid.\ {\bf 55}, 5445 (1997);

\bibitem{mueller}
A.~H.~Mueller, Nucl.\ Phys.\ B {\bf 558}, 285 (1999).

\bibitem{DGLAP}
V.~N.~Gribov and L.~N.~Lipatov, Yad.\ Fiz.\  {\bf 15}, 781 (1972)
[Sov.\ J.\ Nucl.\ Phys.\  {\bf 15}, 438 (1972)];\\
Yu.~L.~Dokshitzer, Sov.\ Phys.\ JETP {\bf 46}, 641 (1977)
[Zh.\ Eksp.\ Teor.\ Fiz.\  {\bf 73}, 1216 (1977)];\\
G.~Altarelli and G.~Parisi, Nucl.\ Phys.\ B {\bf 126}, 298 (1977).

\bibitem{djm2}
A.~Dumitru and J.~Jalilian-Marian,
Phys.\ Rev.\ Lett.\  {\bf 89}, 022301 (2002).

\bibitem{gy_mcl}
M.~Gyulassy and L.~D.~McLerran, Phys.\ Rev.\ C {\bf 56}, 2219 (1997).

\bibitem{FS88}
L.~L.~Frankfurt and M.~I.~Strikman,
Phys.\ Rept.\  {\bf 160}, 235 (1988).

\bibitem{gelis}
F.~Gelis and A.~Peshier, Nucl.\ Phys.\ A {\bf 697}, 879 (2002);\\
F.~Gelis and J.~Jalilian-Marian, Phys.\ Rev.\ D {\bf 67}, 074019 (2003).

\bibitem{Boer}
D.~Boer and A.~Dumitru, Phys.\ Lett.\ B {\bf 556}, 33 (2003);

\bibitem{kazu}
E.~Iancu, K.~Itakura and L.~McLerran, Nucl.\ Phys.\ A {\bf 724}, 181 (2003).

\bibitem{glue}
Y.~V.~Kovchegov and A.~H.~Mueller, Nucl.\ Phys.\ B {\bf 529}, 451 (1998);\\
B.~Z.~Kopeliovich, A.~V.~Tarasov and A.~Sch\"afer,
Phys.\ Rev.\ C {\bf 59}, 1609 (1999);\\
A.~Dumitru and L.~D.~McLerran, Nucl.\ Phys.\ A {\bf 700}, 492 (2002);\\
A.~Dumitru and J.~Jalilian-Marian, Phys.\ Lett.\ B {\bf 547}, 15 (2002);\\
J.~Jalilian-Marian, Y.~Nara and R.~Venugopalan,
Phys.\ Lett.\ B {\bf 577}, 54 (2003);\\
J.~P.~Blaizot, F.~Gelis and R.~Venugopalan,
Nucl.\ Phys.\ A {\bf 743}, 13 (2004).

\bibitem{KL}
D.~Kharzeev, E.~Levin and L.~McLerran, Phys.\ Lett.\ B {\bf 561}, 93 (2003);\\
D.~Kharzeev, E.~Levin and M.~Nardi, Nucl.\ Phys.\ A {\bf 730}, 448 (2004).

\bibitem{KrasVenu}
A.~Krasnitz and R.~Venugopalan,
Phys.\ Rev.\ Lett.\  {\bf 86}, 1717 (2001);\\
A.~Krasnitz, Y.~Nara and R.~Venugopalan,
Phys.\ Rev.\ Lett.\  {\bf 87}, 192302 (2001).

\bibitem{cteq5}
H.~L.~Lai {\it et al.}  [CTEQ Collaboration],
Eur.\ Phys.\ J.\ C {\bf 12}, 375 (2000).

\bibitem{IancuVenu}
E.~Iancu and R.~Venugopalan, arXiv:hep-ph/0303204.

\bibitem{gbw}
K.~Golec-Biernat and M.~Wusthoff, Phys.\ Rev.\ D {\bf 59}, 014017 (1999).

\bibitem{Triantafyllopoulos:2002nz}
D.~N.~Triantafyllopoulos, Nucl.\ Phys.\ B {\bf 648}, 293 (2003).

\bibitem{GRV94p}
M.~Gl\"uck, E.~Reya and A.~Vogt,
Z.\ Phys.\ C {\bf 67}, 433 (1995).

\bibitem{GRV94pi}
M.~Gl\"uck, E.~Reya and A.~Vogt,
Z.\ Phys.\ C {\bf 53}, 651 (1992).

\bibitem{PY}
B.~Andersson, G.~Gustafson, G.~Ingelman and T.~Sj\"ostrand,
Phys.\ Rept.\  {\bf 97}, 31 (1983);\\
T.~Sj\"ostrand, P.~Eden, C.~Friberg, L.~Lonnblad, G.~Miu, S.~Mrenna and
E.~Norrbin, Comput.\ Phys.\ Commun.\  {\bf 135}, 238 (2001).

\bibitem{Sibyll}
R.~S.~Fletcher, T.~K.~Gaisser, P.~Lipari and T.~Stanev,
Phys.\ Rev.\ D {\bf 50}, 5710 (1994);\\
R.~Engel, T.~K.~Gaisser, T.~Stanev and P.~Lipari,
{\it Prepared for 26th International Cosmic Ray Conference (ICRC 99),
  Salt Lake City, Utah, 17-25 Aug 1999}.

\bibitem{qgsjet}
N.~N.~Kalmykov, S.~S.~Ostapchenko and A.~I.~Pavlov,
Nucl.\ Phys.\ Proc.\ Suppl.\  {\bf 52B}, 17 (1997).

\bibitem{BRAHMS}
I.~Arsene {\it et al.}  [BRAHMS Collaboration], arXiv:nucl-ex/0401025.

\bibitem{KL_LHC}
D.~Kharzeev, E.~Levin and M.~Nardi, Nucl.\ Phys.\ A {\bf 747}, 609 (2005).

\bibitem{GaisserBook}
T.~Gaisser, ``Cosmic Rays and Particle Physics'', Cambridge University
Press (1990). 

\bibitem{seneca}
G.~Bossard et al.,  Phys. Rev. {\bf D63}, 054030 (2001);\\
H.~J.~Drescher and G.~Farrar, Phys.\ Rev.\ {\bf D67}, 116001 (2003).

\bibitem{hadrLDF}
H.~J.~Drescher, M.~Bleicher, S.~Soff and H.~St\"ocker,
Astropart.\ Phys.\  {\bf 21}, 87 (2004).

\bibitem{GaisserHalzen}
T.~K.~Gaisser and F.~Halzen,
Phys.\ Rev.\ Lett.\  {\bf 54}, 1754 (1985).

\bibitem{EGS4}
W.~R.\ Nelson, H.~Hirayama and D.~W.~O.~Rogers,
{\em The EGS4 Code System}, report
SLAC-265, Stanford Linear Accelerator Center, 1985.

\bibitem{pylos}
H.~J.~Drescher, arXiv:astro-ph/0411144.

\bibitem{Takeda:2002at}
M.~Takeda et~al., Astropart.\ Phys.\ {\bf 19}, 447 (2003).

\bibitem{Kampert:2004rz}
K.~H.~Kampert {\it et al.}  [The KASCADE Collaboration],
Nucl.\ Phys.\ Proc.\ Suppl.\  {\bf 136}, 273 (2004)
[arXiv:astro-ph/0410559].

\end{thebibliography}
\end{document}